\documentclass[12pt]{article}
\usepackage{amssymb}

\usepackage{amsmath}
\usepackage[unicode,bookmarks,bookmarksopen,bookmarksopenlevel=2,colorlinks,linkcolor=blue,citecolor=green]{hyperref}

\def\comment#1{}
\input{tcilatex}

\begin{document}

\title{The Kupershmidt hydrodynamic chains and lattices}
\author{Maxim V. Pavlov}
\date{}
\maketitle

\begin{abstract}
This paper is devoted to the very important class of hydrodynamic chains
(see \textbf{\cite{Fer+Dav}}, \textbf{\cite{Maks+Egor}}, \textbf{\cite%
{Maks+eps}}) first derived by B. Kupershmidt in \textbf{\cite{Kuper}}, later
re-discovered by M. Blaszak in \textbf{\cite{Blaszak} }(see also \textbf{%
\cite{Manasa}}). An infinite set of local Hamiltonian structures,
hydrodynamic reductions parameterized by the hypergeometric function and
reciprocal transformations for the Kupershmidt hydrodynamic chains are
described.
\end{abstract}

\vspace{1cm}

\textit{In honour of Boris Kupershmidt}

\vspace{1cm}

\tableofcontents

\textit{keywords}: hydrodynamic chain, Hamiltonian structure, reciprocal
transformation, hydrodynamic reduction.


\section{Introduction}

The Kupershmidt hydrodynamic chains ($\beta $ and $\gamma $ are arbitrary
constants)%
\begin{equation}
B_{t}^{k}=B_{x}^{k+1}+\frac{1}{\beta }B^{0}B_{x}^{k}+(k+\gamma
)B^{k}B_{x}^{0}\text{, \ \ \ \ \ \ }k=0,1,2...  \label{1}
\end{equation}%
introduced in \textbf{\cite{Kuper} }recently were re-discovered in \textbf{%
\cite{Blaszak} }for $\beta =1/N$ and $\gamma =1/M$, where $N$ and $M$ are\
integers (see also \textbf{\cite{Manas}}). In this paper we consider three
distinguished features of these hydrodynamic chains:

\textbf{1}. new explicit hydrodynamic reductions determined by the
hypergeometric function;

\textbf{2}. infinitely many local and nonlocal Hamiltonian structures;

\textbf{3}. reciprocal transformations connecting the Kupershmidt
hydrodynamic chains (\textbf{\ref{1}}) with the distinct parameters $\beta $
to each other.

Also we discuss in details the integrability of these hydrodynamic chains by
the generalized hodograph method and related 2+1 quasilinear systems.

The paper is organized in the following order. In the second section the
Gibbons equation describing a deformation of the Riemann mapping associated
with the Kupershmidt hydrodynamic chain and deformations of the Riemann
surfaces associated with corresponding hydrodynamic reductions is derived.
In the third section we prove that the Gibbons--Tsarev system is common for
all Kupershmidt hydrodynamic chains (irrespective of the distinct parameters 
$\beta $) as well as for the Benney hydrodynamic chain. In the fourth
section infinitely many hydrodynamic reductions are found. In the fifth
section infinitely many local Hamiltonian structures are constructed. Also
nonlocal Hamiltonian structure connected with a metric of constant curvature
is presented. In the sixth section the extended Kupershmidt hydrodynamic
chains are constructed. In the seventh section linear transformations of
independent variables preserving the Kupershmidt hydrodynamic chains are
described. In the eighth section reciprocal transformations preserving the
Kupershmidt hydrodynamic chains are found. In the ninth section the ideal
gas dynamics as simplest two component hydrodynamic reduction of the
Kupershmidt hydrodynamic chains is considered. In the tenth section the
Kupershmidt hydrodynamic chains determined by special values of the
parameter $\beta =N$ are derived. In the eleventh section multi-parametric
solutions given by the generalized hodograph method are obtained for 2+1
quasilinear equations associated with the Kupershmidt hydrodynamic lattice.

\section{The Gibbons equation and explicit hydrodynamic reductions}

The Kupershmidt hydrodynamic chain (\textbf{\ref{1}}) is a \textit{linear}
hydrodynamic chain with respect to discrete variable $k$ (i.e. the r.h.s. of
(\textbf{\ref{1}}) is a \textit{linear} function with respect to the
discrete variable $k$). Let us seek the moment decomposition (see \textbf{%
\cite{Maks+Hamch}}) in the form%
\begin{equation}
B^{k}=\frac{1}{k+\gamma }\underset{i=1}{\overset{N}{\sum }}\varepsilon
_{i}(b^{i})^{\beta (k+\gamma )}\text{, \ \ \ \ \ }\gamma \neq 0,-1,-2,...
\label{one}
\end{equation}%
Then (\textbf{\ref{1}}) reduces to the hydrodynamic type system%
\begin{equation}
b_{t}^{i}=\partial _{x}\left[ \frac{(b^{i})^{\beta +1}}{\beta +1}+\frac{B^{0}%
}{\beta }b^{i}\right] \text{, \ \ \ \ \ \ }i=1,2,...,N.  \label{two}
\end{equation}%
The generating function of conservation laws of this hydrodynamic reduction%
\begin{equation}
p_{t}=\partial _{x}\left( \frac{p^{\beta +1}}{\beta +1}+\frac{B^{0}}{\beta }%
p\right)  \label{tri}
\end{equation}%
can be obtained (see \textbf{\cite{Maks+algebr}}) by the replacement $%
b^{i}\rightarrow p$. Simultaneously, (\textbf{\ref{tri}}) is the generating
function of conservation laws for the Kupershmidt hydrodynamic chains (%
\textbf{\ref{1}}).

\textbf{Theorem 1}: \textit{The Gibbons equation}%
\begin{equation}
\lambda _{t}-\left( p^{\beta }+\frac{B^{0}}{\beta }\right) \lambda _{x}=%
\frac{\partial \lambda }{\partial p}\left[ p_{t}-\partial _{x}\left( \frac{%
p^{\beta +1}}{\beta +1}+\frac{B^{0}}{\beta }p\right) \right]  \label{8}
\end{equation}%
\textit{describes a deformation of the Riemann mapping} $\lambda (\mathbf{b}%
;p)$ \textit{determined by the series}%
\begin{equation}
\lambda =q^{1-\gamma }+(1-\gamma )\underset{k=0}{\overset{\infty }{\sum }}%
\frac{B^{k}}{q^{k+\gamma }}\text{, \ \ \ \ \ }\lambda \rightarrow \infty ,
\label{rim}
\end{equation}%
\textit{where }$q=p^{\beta }$\textit{\ and the coefficients} $B^{k}(\mathbf{b%
};p)$ \textit{satisfy} (\textbf{\ref{1}}).

\textbf{Proof}: If (\textbf{\ref{tri}}) and (\textbf{\ref{two}}) are
consistent, then%
\begin{equation}
\partial _{i}q=q\frac{\partial _{i}B^{0}}{q-u^{i}}\left[ \sum \frac{%
u^{n}\partial _{n}B^{0}}{q-u^{n}}-1\right] ^{-1},  \label{comp}
\end{equation}%
where $u^{i}=(b^{i})^{\beta }$ and $\partial _{i}=\partial /\partial u^{i}$.
Since $\partial _{i}q=-\partial _{i}\lambda /\partial _{q}\lambda $ (this is
a consequence of the consistency (\textbf{\ref{two}}) with (\textbf{\ref{8}}%
)), then the equation of the Riemann surface can be found in quadratures%
\begin{equation}
d\lambda =q^{-\gamma }\left[ \sum \frac{\varepsilon _{n}(u^{n})^{\gamma }}{%
q-u^{n}}-1\right] dq-q^{1-\gamma }\sum \frac{\varepsilon _{n}(u^{n})^{\gamma
-1}}{q-u^{n}}du^{n}.  \label{dif}
\end{equation}%
Thus, the equation of the Riemann surface%
\begin{equation}
\lambda =q^{1-\gamma }+(1-\gamma )\sum \varepsilon _{n}\left( \frac{u^{n}}{q}%
\right) ^{\gamma }F\left( 1,\gamma ,\gamma +1,\frac{u^{n}}{q}\right)
\label{giper}
\end{equation}%
connected with (\textbf{\ref{two}}) is parameterized by the hypergeometric
function $_{2}F_{1}(a,b,c,z)$. Then, the substitution (\textbf{\ref{one}})
in the above formula leads to the equation of the Riemann mapping (\textbf{%
\ref{rim}}) for (\textbf{\ref{1}}). The theorem is proved.

If $\gamma =0$, then the Kupershmidt hydrodynamic chain%
\begin{equation}
B_{t}^{k}=B_{x}^{k+1}+\frac{1}{\beta }B^{0}B_{x}^{k}+kB^{k}B_{x}^{0}\text{,
\ \ \ \ \ \ }k=0,1,2...  \label{can}
\end{equation}%
is connected with the equation of the Riemann mapping (\textbf{\ref{rim}})%
\begin{equation}
\lambda =q+\underset{k=0}{\overset{\infty }{\sum }}\frac{B^{k}}{q^{k}}\text{%
, \ \ \ \ \ }\lambda \rightarrow \infty .  \label{kan}
\end{equation}

\textbf{Definition 1}: The Kupershmidt hydrodynamic chain (\textbf{\ref{can}}%
) is said to be written in the \textit{canonical} form.

Let us introduce the sub-index $\gamma $ for the moments $B_{(\gamma )}^{k}$%
, which satisfy the Kupershmidt hydrodynamic chain (\textbf{\ref{1}})%
\begin{equation}
\partial _{t}B_{(\gamma )}^{k}=\partial _{x}B_{(\gamma )}^{k+1}+\frac{1}{%
\beta }B^{0}\partial _{x}B_{(\gamma )}^{k}+(k+\gamma )B_{(\gamma
)}^{k}B_{x}^{0}\text{, \ \ \ \ \ \ }k=0,1,2...,  \label{fuk}
\end{equation}%
where $B_{(\gamma )}^{0}\equiv B^{0}$. The invertible \textit{polynomial}
transformations $B_{(\gamma )}^{k}=B_{(\gamma )}^{k}(B^{0},B^{1},...,B^{k})$
can be obtained by the comparison (\textbf{\ref{rim}}) with (\textbf{\ref%
{kan}})%
\begin{equation}
\lambda =q+\underset{k=0}{\overset{\infty }{\sum }}\frac{B^{k}}{q^{k}}=\left[
q^{1-\gamma }+(1-\gamma )\underset{k=0}{\overset{\infty }{\sum }}\frac{%
B_{(\gamma )}^{k}}{q^{k+\gamma }}\right] ^{\frac{1}{1-\gamma }}.  \label{ser}
\end{equation}%
For instance,%
\begin{equation}
B_{(\gamma )}^{1}=B^{1}-\frac{\gamma }{2}(B^{0})^{2}\text{, \ \ \ \ \ }%
B_{(\gamma )}^{2}=B^{2}-\gamma B^{0}B^{1}+\frac{\gamma (\gamma +1)}{6}%
(B^{0})^{3},...  \label{inver}
\end{equation}

If $\gamma =1$, then the equation of the Riemann mapping (\textbf{\ref{rim}}%
) reduces to%
\begin{equation}
\lambda =\ln q+\underset{k=0}{\overset{\infty }{\sum }}\frac{B_{(1)}^{k}}{%
q^{k+1}}\text{ \ \ \ \ }\Leftrightarrow \text{ \ \ \ \ }\lambda =q\exp
\left( \underset{k=0}{\overset{\infty }{\sum }}\frac{B_{(1)}^{k}}{q^{k+1}}%
\right) .  \label{log}
\end{equation}%
These above formulas are equivalent up to scaling $\lambda \rightarrow \exp
\lambda $; since the Gibbons equation is a \textit{linear} equation with
respect to $\lambda $, any scaling $\lambda \rightarrow \tilde{\lambda}%
(\lambda )$ is admissible. Thus, the Kupershmidt hydrodynamic chains (%
\textbf{\ref{can}}) and (\textbf{\ref{fuk}}) are equivalent under the above
invertible transformations (see \textbf{\cite{Kuper}}).

\textbf{Remark}: The rational hydrodynamic reductions of the Kupershmidt
hydrodynamic chain (\textbf{\ref{1}}) are found in \textbf{\cite{Blaszak}}
(two-component hydrodynamic reductions are described in \textbf{\cite{Manas}}%
). The above hydrodynamic reductions (\textbf{\ref{two}}) connected with the
equation of the Riemann surface (\textbf{\ref{giper}}) are new, but still
are not the most general. Most complicated (known at this moment)
hydrodynamic reductions are considered below. Some of them are connected
with the Hamiltonian structures of the Kupershmidt hydrodynamic chain (%
\textbf{\ref{1}}).

If $\gamma =0,-1,-2,...$, some of the moments $B_{(\gamma )}^{k}$ simplify.
More precisely, the moment decomposition for these exceptional cases is
given by (for $\gamma =-K$, $K=0,1,2,...$)%
\begin{equation}
B_{(-K)}^{k}|_{k\neq K}=\frac{1}{k-K}\underset{i=1}{\overset{N}{\sum }}%
\varepsilon _{i}(u^{i})^{k-K}\text{, \ \ \ \ \ }B_{(-K)}^{K}=\underset{i=1}{%
\overset{N}{\sum }}\varepsilon _{i}\ln u^{i}\text{.}  \label{exep}
\end{equation}%
This degenerate case leads to the constraint $\Sigma \varepsilon _{k}=0$.
Let us emphasize again that this constraint $\Sigma \varepsilon _{k}=0$
exists for the exceptional cases $\gamma =0,-1,-2,...$ only (in all other
cases it is still possible but an unessential restriction).

The Gibbons equation (\textbf{\ref{8}}) (see \textbf{\cite{Gibbons}}, 
\textbf{\cite{Maks+algebr}}) has three distinguished features.

\textbf{1}. If $\lambda =\limfunc{const}$, then $\partial \lambda /\partial
p\neq 0$ and (\textbf{\ref{8}}) reduces to (\textbf{\ref{tri}}). Taking into
account $q=p^{\beta }$ and substituting the inverse (the B\"{u}%
rmann--Lagrange) series%
\begin{equation*}
q=\lambda -\underset{k=0}{\overset{\infty }{\sum }}\frac{Q^{k}(\mathbf{B})}{%
\lambda ^{k}}\text{, \ \ \ \ \ }q\rightarrow \infty
\end{equation*}%
in (\textbf{\ref{tri}}) and (\textbf{\ref{kan}}), one can obtain an infinite
series of polynomial conservation laws (see, for instance, the sections 
\textbf{5} and \textbf{7}).

\textbf{2}. If $p=\limfunc{const}$, then (\textbf{\ref{8}}) reduces to the
kinetic equation written in the Lax form (see \textbf{\cite{Gibbons}}, 
\textbf{\cite{Zakh}})%
\begin{equation*}
\lambda _{t}=\{\lambda ,\mathbf{\hat{H}}\}\equiv \frac{\partial \mathbf{%
\hat{H}}}{\partial p}\frac{\partial \lambda }{\partial x}-\frac{\partial 
\mathbf{\hat{H}}}{\partial x}\frac{\partial \lambda }{\partial p},
\end{equation*}%
where $\mathbf{\hat{H}}=p^{\beta +1}/(\beta +1)+B^{0}p/\beta $ (cf. the flux
of (\textbf{\ref{tri}})). This Lax formulation can be derived as a
dispersionless limit from $R-$matrix approach (see \textbf{\cite{Blaszak}})
for $\beta =1/N$%
\begin{equation*}
\lambda _{t}=\{\lambda ,\mathbf{\hat{H}}\}\equiv q^{1-N}\left[ \frac{%
\partial \mathbf{\hat{H}}}{\partial q}\frac{\partial \lambda }{\partial x}-%
\frac{\partial \mathbf{\hat{H}}}{\partial x}\frac{\partial \lambda }{%
\partial q}\right] ,
\end{equation*}%
where $\mathbf{\hat{H}}=q^{N+1}/(N+1)+B^{0}q^{N}$.

\textbf{3}. If $\partial \lambda /\partial p=0$, then r.h.s. of (\textbf{\ref%
{8}}) is vanished, and the hydrodynamic reductions (cf. (\textbf{\ref{two}}))%
\begin{equation}
b_{t}^{i}=\partial _{x}\left[ \frac{(b^{i})^{\beta +1}}{\beta +1}+\frac{%
B^{0}(\mathbf{b})}{\beta }b^{i}\right] \text{, \ \ \ \ \ \ }i=1,2,...,N,
\label{3}
\end{equation}%
can be written in the Riemann invariants $r^{k}=\lambda |_{\partial \lambda
/\partial p=0}$%
\begin{equation}
r_{t}^{i}=\left[ (p^{i})^{\beta }+\frac{B^{0}(\mathbf{r})}{\beta }\right]
r_{x}^{i}\text{, \ \ \ \ \ \ }i=1,2,...,N,  \label{2}
\end{equation}%
where $B^{0}$ is a solution of some nonlinear PDE system describing $N$
component hydrodynamic reductions parameterized by $N$ arbitrary functions
of a single variable (see \textbf{\cite{Gib+Tsar}}), which we call the
Gibbons--Tsarev system.

\section{\textit{Universality} of the Gibbons--Tsarev system}

The consistency of the generating function of conservation laws (\textbf{\ref%
{tri}}) with the hydrodynamic type system (\textbf{\ref{2}}) yields (cf. (%
\textbf{\ref{comp}}))%
\begin{equation*}
\partial _{i}q=q\frac{\partial _{i}B^{0}}{q^{i}-q},
\end{equation*}%
where $q^{i}=(p^{i})^{\beta }$ and $\partial _{i}\equiv \partial /\partial
r^{i}$. The compatibility conditions $\partial _{k}(\partial _{i}q)=\partial
_{i}(\partial _{k}q)$ yield the famous Gibbons--Tsarev system (see \textbf{%
\cite{Gib+Tsar}})%
\begin{equation}
\partial _{i}\mu ^{k}=\frac{\partial _{i}A^{0}}{\mu ^{i}-\mu ^{k}}\text{, \
\ \ \ \ \ \ }\partial _{ik}A^{0}=2\frac{\partial _{i}A^{0}\partial _{k}A^{0}%
}{(\mu ^{i}-\mu ^{k})^{2}}\text{, \ \ \ \ \ \ \ }i\neq k,  \label{gt}
\end{equation}%
where $\mu ^{i}=q^{i}+B^{0}$ and the potential $A^{0}$ can be reconstructed
in quadratures from $\partial _{i}A^{0}=q^{i}\partial _{i}B^{0}$. Since the
Gibbons--Tsarev system was derived for a description of hydrodynamic
reductions of the Benney hydrodynamic chain (see \textbf{\cite{Benney}})%
\begin{equation}
A_{t}^{k}=A_{x}^{k+1}+kA^{k-1}A_{x}^{0}\text{, \ \ \ \ \ \ }k=0,1,2,...,
\label{bm}
\end{equation}%
then hydrodynamic reductions of the different (\textit{any} values of the
parameter $\beta $) Kupershmidt hydrodynamic chains are the \textbf{same} as
hydrodynamic reductions of the Benney hydrodynamic chain up to the
aforementioned transformations $\partial _{i}A^{0}=q^{i}\partial _{i}B^{0}$
and $\mu ^{i}=q^{i}+B^{0}$.

\section{Explicit hydrodynamic reductions}

The Gibbons--Tsarev system (\textbf{\ref{gt}}) is integrable but a general
solution is not found yet. However, in this section we shall be able to
present the method allowing to find infinitely many explicit hydrodynamic
reductions without solving the Gibbons--Tsarev system (\textbf{\ref{gt}}).

In the previous section we mentioned that any hydrodynamic reduction (%
\textbf{\ref{2}}) of the Kupershmidt hydrodynamic chain (\textbf{\ref{fuk}})
can be written in the form (\textbf{\ref{3}}). However, this is not a unique
choice. Following \textbf{\cite{Gib+Tsar}} an arbitrary hydrodynamic
reduction (\textbf{\ref{2}}) can be written via the first $N$ moments $%
B_{(\gamma )}^{k}$%
\begin{eqnarray*}
\partial _{t}B_{(\gamma )}^{k} &=&\partial _{x}B_{(\gamma )}^{k+1}+\frac{1}{%
\beta }B^{0}\partial _{x}B_{(\gamma )}^{k}+kB_{(\gamma )}^{k}B_{x}^{0}\text{%
, \ \ \ \ \ \ }k=0\text{, }1\text{, }2\text{, ..., }N-2, \\
&& \\
\partial _{t}B_{(\gamma )}^{N-1} &=&\partial _{x}B_{(\gamma )}^{N}(\mathbf{B}%
)+\frac{1}{\beta }B^{0}\partial _{x}B_{(\gamma )}^{N-1}+(N-1)B_{(\gamma
)}^{N-1}B_{x}^{0}\text{,}
\end{eqnarray*}%
where $B_{(\gamma )}^{N}(\mathbf{B})$ is a some function of the first $N$
moments $B_{(\gamma )}^{k}$ compatible with the generating function of
conservation laws (\textbf{\ref{tri}}). This hydrodynamic reduction
significantly simplifies under the constraints $B^{N}=B^{N+1}=...=0$ (see 
\textbf{\cite{Blaszak}}). The similar constraints $B_{(\gamma
)}^{N}=B_{(\gamma )}^{N+1}=...=0$ are compatible with the Kupershmidt
hydrodynamic chain (see \textbf{\cite{Blaszak}} again) written in the form (%
\textbf{\ref{fuk}}). The corresponding hydrodynamic type system%
\begin{eqnarray}
\partial _{t}B_{(\gamma )}^{n} &=&\partial _{x}B_{(\gamma )}^{n+1}+\frac{1}{%
\beta }B^{0}\partial _{x}B_{(\gamma )}^{n}+(n+\gamma )B_{(\gamma
)}^{n}B_{x}^{0}\text{, \ \ \ }n=0\text{, }1\text{, }2\text{, ..., }N-2, 
\notag \\
&&  \label{e} \\
\partial _{t}B_{(\gamma )}^{N-1} &=&\frac{1}{\beta }B^{0}\partial
_{x}B_{(\gamma )}^{N-1}+(N-1+\gamma )B_{(\gamma )}^{N-1}B_{x}^{0}\text{.} 
\notag
\end{eqnarray}%
is connected with the equation of the Riemann surface (cf. (\textbf{\ref{rim}%
}))%
\begin{equation}
\lambda =q^{1-\gamma }+(1-\gamma )\underset{k=0}{\overset{N-1}{\sum }}\frac{%
B_{(\gamma )}^{k}}{q^{k+\gamma }}.  \label{mop}
\end{equation}

\textbf{Remark}: The Kupershmidt hydrodynamic chains are equivalent under
the invertible transformations $B_{(\gamma )}^{k}=B_{(\gamma
)}^{k}(B^{0},B^{1},...,B^{k})$, but the corresponding reductions (\textbf{%
\ref{e}}) are different for an arbitrary values of parameter $\gamma $. It
is obvious if to take into account (\textbf{\ref{mop}}) (cf. (\textbf{\ref%
{ser}}))%
\begin{equation*}
q+\underset{k=0}{\overset{N-1}{\sum }}\frac{B^{k}}{q^{k}}\neq \left[
q^{1-\gamma }+(1-\gamma )\underset{k=0}{\overset{N-1}{\sum }}\frac{%
B_{(\gamma )}^{k}}{q^{k+\gamma }}\right] ^{\frac{1}{1-\gamma }}.
\end{equation*}

An arbitrary ($K+M$ component) hydrodynamic reduction (\textbf{\ref{2}}) can
be written in the \textit{mixed} form%
\begin{eqnarray*}
\partial _{t}B_{(\gamma )}^{k} &=&\partial _{x}B_{(\gamma )}^{k+1}+\frac{1}{%
\beta }B^{0}\partial _{x}B_{(\gamma )}^{k}+(k+\gamma )B_{(\gamma
)}^{k}B_{x}^{0}\text{, \ \ \ }k=0\text{, }1\text{, }2\text{, ..., }K-2, \\
&& \\
\partial _{t}B_{(\gamma )}^{K-1} &=&\partial _{x}B_{(\gamma )}^{K}(\mathbf{%
B,b})+\frac{1}{\beta }B^{0}\partial _{x}B_{(\gamma )}^{K-1}+(K-1+\gamma
)B_{(\gamma )}^{K-1}B_{x}^{0}\text{,} \\
&& \\
b_{t}^{m} &=&\partial _{x}\left( \frac{(b^{m})^{\beta +1}}{\beta +1}+\frac{%
B^{0}}{\beta }b^{m}\right) \text{, \ \ \ \ \ \ \ }m=1\text{, }2\text{, ..., }%
M,
\end{eqnarray*}%
where $B^{K}(\mathbf{B}$, $\mathbf{b})$ is a some function (of the punctures 
$b^{m}$ and the first moments $B^{k}$), which must be compatible with the
generating function of conservation laws (\textbf{\ref{tri}}). We avoid a
derivation of a nonlinear PDE system on function $B_{(\gamma )}^{K}(\mathbf{%
B,b})$, because the corresponding system is equivalent the Gibbons--Tsarev
system (see the previous section). However, one particular solution can be
found in the explicit form without this complicated analysis.

\textbf{Main result of this section}: a most general \textit{explicit}
hydrodynamic reduction found at this moment%
\begin{eqnarray}
\partial _{t}B_{(\gamma )}^{k} &=&\partial _{x}B_{(\gamma )}^{k+1}+\frac{1}{%
\beta }B^{0}\partial _{x}B_{(\gamma )}^{k}+(k+\gamma )B_{(\gamma
)}^{k}B_{x}^{0}\text{, \ \ \ }k=0,1,...,K-2,  \notag \\
&&  \notag \\
\partial _{t}B_{(\gamma )}^{K-1} &=&\partial _{x}B_{(\gamma )}^{K}(\mathbf{b}%
)+\frac{1}{\beta }B^{0}\partial _{x}B_{(\gamma )}^{K-1}+(K-1+\gamma
)B_{(\gamma )}^{K-1}B_{x}^{0}\text{,}  \label{t} \\
&&  \notag \\
b_{t}^{k} &=&\partial _{x}\left( \frac{(b^{k})^{\beta +1}}{\beta +1}+\frac{%
B^{0}}{\beta }b^{k}\right) \text{, \ \ \ \ \ \ \ }k=1,2,...,M,  \notag
\end{eqnarray}%
is connected with the equation of the Riemann surface (cf. (\textbf{\ref%
{giper}}) and (\textbf{\ref{mop}}))%
\begin{equation}
\lambda =\frac{q^{1-\gamma }}{1-\gamma }+\underset{k=0}{\overset{K-1}{\sum }}%
\frac{B_{(\gamma )}^{k}}{q^{k+\gamma }}+\underset{m=1}{\overset{M}{\sum }}%
\varepsilon _{m}\left( \frac{u^{m}}{q}\right) ^{K+\gamma }F\left( 1,K+\gamma
,K+1+\gamma ,\frac{u^{m}}{q}\right) ,  \label{r}
\end{equation}%
while all higher moments ($\gamma \neq -K,$ $-K-1,$ $-K-2,...$; see (\textbf{%
\ref{exep}})) are given by (see (\textbf{\ref{one}}))%
\begin{equation}
B_{(\gamma )}^{K+n}=\frac{1}{K+n+\gamma }\underset{m=1}{\overset{M}{\sum }}%
\varepsilon _{m}(u^{m})^{K+n+\gamma }\text{, \ \ \ \ \ \ }n=0,1,2,...
\label{k}
\end{equation}%
The equation of the Riemann surface (\textbf{\ref{r}}) can be simplified for 
$\gamma =L_{1}/L_{2}$ (where $L_{1}$ and $L_{2}$ and integers), because a
hypergeometric function degenerates in elementary functions. If $\gamma =-K,$
$-K-1,$ $-K-2,...$, then the extra constraint $\Sigma \varepsilon _{m}=0$
appears.

If $\gamma =1$, then the above hydrodynamic reductions (\textbf{\ref{t}})
become \textit{homogeneous} hydrodynamic type systems; if, for instance, $%
\gamma =1$ and $K=0$, then so-called the Schwarz-Christoffel type reduction
(cf. \textbf{\cite{Gib+Tsar}}) is determined by%
\begin{equation*}
\lambda =q\underset{k=1}{\overset{N}{\prod }}\left( 1-\frac{u^{k}}{q}\right)
^{-\varepsilon _{k}}.
\end{equation*}

Indeed, the hydrodynamic type system (\textbf{\ref{t}}) is a hydrodynamic
reduction of the Kupershmidt hydrodynamic chain (\textbf{\ref{fuk}}). It is
easy to verify taking into account that (\textbf{\ref{t}}) is compatible
with more deep reduction (\textbf{\ref{one}}). It means that the
hydrodynamic reduction (\textbf{\ref{t}}) can be obtained from (\textbf{\ref%
{fuk}}) by the moment decomposition (\textbf{\ref{k}}) applied to the higher
($n\geqslant K$) moments $B_{(\gamma )}^{n}$ only. Then the first ($n<K$)
moments $B_{(\gamma )}^{n}$ are natural field variables as well as the
punctures $b^{k}$.

All other reductions can be obtained by different parametric and functional
degenerations. The substitution of the Taylor series (we remember that $%
q=p^{\beta }$)%
\begin{equation*}
q=\underset{k=0}{\overset{\infty }{\sum }}q^{k}\lambda ^{k}\text{, \ \ \ \ \
\ }\lambda \rightarrow 0
\end{equation*}%
in (\textbf{\ref{tri}}) yields%
\begin{equation}
q_{t}^{k}=\frac{B^{0}}{\beta }q_{x}^{k}+q^{k}B_{x}^{0}+\underset{m=0}{%
\overset{k}{\sum }}q^{k-m}q_{x}^{m}\text{, \ \ \ \ }k=0,1,2,...  \label{m}
\end{equation}%
Let us consider $M$ such series \textit{truncated} up to some numbers $M_{k}$%
\begin{equation*}
\partial _{t}q_{(k)}^{m}=\frac{B^{0}}{\beta }\partial
_{x}q_{(k)}^{m}+q_{(k)}^{m}B_{x}^{0}+\underset{n=0}{\overset{m}{\sum }}%
q_{(k)}^{m-n}\partial _{x}q_{(k)}^{n}\text{, \ \ \ \ }m=0,1,...,M_{k}\text{,
\ \ \ \ }k=0,1,...,M.
\end{equation*}%
If $M_{k}=0$, this is the general case (\textbf{\ref{t}}). Exceptional cases
($M_{k}\neq 0$) can be obtained by the \textit{merging} of neighbouring
singular points $q_{(k)}^{0}\rightarrow q_{(k+1)}^{0}$. For instance, if $%
M_{k}=1$ and $K=0$, the Kupershmidt hydrodynamic chain (\textbf{\ref{fuk}})
has a one parametric family of the Zakharov hydrodynamic reductions (the
parameter $\gamma $ is arbitrary for the each fixed parameter $\beta $)%
\begin{equation*}
b_{t}^{i}=\partial _{x}\left[ \frac{(b^{i})^{\beta +1}}{\beta +1}+\frac{B^{0}%
}{\beta }b^{i}\right] \text{, \ \ \ \ \ }c_{t}^{i}=\partial _{x}\left[
\left( (b^{i})^{\beta }+\frac{B^{0}}{\beta }\right) c^{i}\right] \text{, \ \
\ \ \ }i=1,2,...,N
\end{equation*}%
connected with the equation of the Riemann surface%
\begin{equation*}
\lambda =q^{1-\gamma }\left[ 1+(1-\gamma )\underset{k=1}{\overset{N}{\sum }}%
\frac{(b^{k})^{\beta \gamma -1}c^{k}}{q-(b^{k})^{\beta }}\right] \text{, \ \
\ \ \ }\gamma \neq 1,
\end{equation*}%
where%
\begin{equation*}
B_{(\gamma )}^{n}=\underset{k=1}{\overset{N}{\sum }}(b^{k})^{\beta (n+\gamma
)-1}c^{k}\text{, \ \ \ \ \ }n=0,1,2,...
\end{equation*}%
If $\gamma =1$, then%
\begin{equation*}
\lambda =\ln q+\underset{k=1}{\overset{N}{\sum }}\frac{(b^{k})^{\beta
-1}c^{k}}{q-(b^{k})^{\beta }}\text{.}
\end{equation*}

Let us take for simplicity just one series (\textbf{\ref{m}}) (i.e. $M=1$)
truncated by the number $L$ (this is exactly the case considered in \textbf{%
\cite{Blaszak}}). Then hydrodynamic reduction (\textbf{\ref{t}}) reduces to%
\begin{eqnarray*}
\partial _{t}B_{(\gamma )}^{k} &=&\partial _{x}B_{(\gamma )}^{k+1}+\frac{1}{%
\beta }B^{0}\partial _{x}B_{(\gamma )}^{k}+(k+\gamma )B_{(\gamma
)}^{k}B_{x}^{0}\text{, \ \ \ }k=0,1,...,K-2, \\
&& \\
\partial _{t}B_{(\gamma )}^{K-1} &=&\partial _{x}B_{(\gamma )}^{K}(\mathbf{q}%
)+\frac{1}{\beta }B^{0}\partial _{x}B_{(\gamma )}^{K-1}+(K-1+\gamma
)B_{(\gamma )}^{K-1}B_{x}^{0}\text{,} \\
&& \\
q_{t}^{k} &=&\frac{B^{0}}{\beta }q_{x}^{k}+q^{k}B_{x}^{0}+\underset{m=0}{%
\overset{k}{\sum }}q^{k-m}q_{x}^{m}\text{, \ \ \ \ \ \ \ }k=0,1,...,L-1,
\end{eqnarray*}%
where the function $B_{(\gamma )}^{K}(\mathbf{q})$ can be found from the
linear PDE system%
\begin{equation*}
q\frac{\partial f}{\partial q}+\underset{k=0}{\overset{L-1}{\sum }}q^{k}%
\frac{\partial f}{\partial q^{k}}=0\text{, \ \ \ \ \ }q\frac{\partial f}{%
\partial q^{k}}-\underset{m=0}{\overset{L-k-1}{\sum }}q^{m}\frac{\partial f}{%
\partial q^{m+k}}=q^{1-K-\gamma }\frac{\partial B_{(\gamma )}^{K}(\mathbf{q})%
}{\partial q^{k}}\text{, \ \ }k=0,1,...,L-1.
\end{equation*}%
In this case the equation of the Riemann surface is given by%
\begin{equation*}
\lambda =\frac{q^{1-\gamma }}{1-\gamma }+\underset{k=0}{\overset{K-1}{\sum }}%
\frac{B_{(\gamma )}^{k}}{q^{k+\gamma }}+f(q^{0},q^{1},...,q^{L-1};q).
\end{equation*}

\section{The Hamiltonian structures}

The theory of Hamiltonian operators for hydrodynamic chains starts from 
\textbf{\cite{KM}}, where the first such Hamiltonian structure was derived,
and from \textbf{\cite{Dorfman}}, where the Jacobi identity for local
Poisson brackets was presented. If such a local Poisson bracket is written
in the Liouville coordinates $B^{k}$%
\begin{equation*}
\{B^{k},B^{n}\}=[\mathcal{W}^{kn}(\mathbf{B})\partial _{x}+\partial _{x}%
\mathcal{W}^{nk}(\mathbf{B})]\delta (x-x^{\prime }),
\end{equation*}%
then the Jacobi identity reduced to the most compact form (see \textbf{\cite%
{Maks+Puas}})%
\begin{eqnarray}
(\mathcal{W}^{ik}+\mathcal{W}^{ki})\partial _{k}\mathcal{W}^{nj} &=&(%
\mathcal{W}^{jk}+\mathcal{W}^{kj})\partial _{k}\mathcal{W}^{ni},  \notag \\
&&  \label{jac} \\
\partial _{n}\mathcal{W}^{ij}\partial _{m}\mathcal{W}^{kn} &=&\partial _{n}%
\mathcal{W}^{kj}\partial _{m}\mathcal{W}^{in}.  \notag
\end{eqnarray}

The very important class of the Poisson brackets determined by $\mathcal{W}%
^{kn}=\mathcal{W}_{(L)}^{kn}(B^{0},B^{1},...,B^{k+n-L})$, which we call the $%
L-$brackets, can be completely described (see \textbf{\cite{Maks+Puas}}).

In this paper we deal with $L-$brackets.

\textbf{Main statement of this section}: The Kupershmidt hydrodynamic chains
(\textbf{\ref{1}}) has \textbf{infinitely} many \textit{local} Hamiltonian
structures enumerated by the index $L$ for $L\geqslant 0$ and infinitely
many nonlocal Hamiltonian structures if $L<0$.

If $L=0$. The important example of such a Poisson bracket is the Kupershmidt
Poisson brackets (see \textbf{\cite{Kuper}})%
\begin{equation}
\{B^{k}\text{, }B^{n}\}=[(\beta k+\gamma )B^{k+n}\partial _{x}+(\beta
n+\gamma )\partial _{x}B^{k+n}]\delta (x-x^{\prime })  \label{x}
\end{equation}%
connected with the Kupershmidt hydrodynamic chains (see below).

These Poisson brackets have the momentum $B^{0}$ only.

If $L=1$. The famous example of such a Poisson bracket is the
Kupershmidt--Manin bracket found in \textbf{\cite{KM}} for the Benney
hydrodynamic chain (\textbf{\ref{bm}})%
\begin{equation*}
\{B^{k},B^{n}\}=[kB^{k+n-1}\partial _{x}+n\partial _{x}B^{k+n-1}]\delta
(x-x^{\prime }).
\end{equation*}%
This Poisson bracket has the momentum $B^{1}$ and the Casimir (annihilator) $%
B^{0}$.

If $L\geqslant 1$, then corresponding Poisson brackets have $L$ Casimirs.
Since the Hamiltonian is $\mathbf{\bar{H}}_{L+1}\mathbf{=}\int \mathbf{H}%
_{L+1}(B^{0},B^{1},B^{2},...,B^{L+1})dx$, then the momentum can be chosen as 
$\mathbf{\bar{H}}_{L}\mathbf{=}\int B^{L}dx$. The Casimirs can be chosen as $%
\mathbf{\bar{H}}_{k}\mathbf{=}\int B^{k}dx$ ($k=0,1,2,...,L-1$). Then the 
\textit{auxiliary} (natural) restrictions (``normalization'') are%
\begin{eqnarray}
B^{k} &=&\mathcal{W}_{(L)}^{Lk}(B^{0},B^{1},...,B^{k})\text{, \ \ \ \ \ \ }%
k=0,1,2,...,  \notag \\
&&  \notag \\
0 &=&\mathcal{W}_{(L)}^{sk}(B^{0},B^{1},...,B^{k+s-L})\text{, \ \ \ \ \ }%
0\leqslant s<L\text{, \ \ \ \ \ \ }k\geqslant L-s.  \notag \\
&&  \notag \\
\mathcal{W}_{(L)}^{kn} &=&\mathcal{\bar{W}}_{(L)}^{kn}=\limfunc{const}\text{%
, \ \ \ \ \ \ }k=0,1,2,...,L-1\text{, \ \ \ \ \ }0\leqslant n\leqslant L-1-k.
\label{y}
\end{eqnarray}

Thus, an arbitrary hydrodynamic chain determined by such $L-$bracket has at
least $L+2$ conservation laws (for an arbitrary Hamiltonian density $\mathbf{%
H}_{L+1}$), where the first $L$ conservation laws of the Casimirs are%
\begin{equation*}
B_{t}^{k}=\partial _{x}\left( \underset{n=0}{\overset{L-k-1}{\sum }}(%
\mathcal{\bar{W}}_{(L)}^{kn}+\mathcal{\bar{W}}_{(L)}^{nk})\frac{\partial 
\mathbf{H}_{L+1}}{\partial B^{n}}+\underset{n=L-k}{\overset{L+1}{\sum }}%
\mathcal{W}_{(L)}^{nk}\frac{\partial \mathbf{H}_{L+1}}{\partial B^{n}}%
\right) \text{, \ \ \ \ }k=0,1,2,...,L-1.
\end{equation*}%
The conservation law of the momentum is%
\begin{equation*}
B_{t}^{L}=\partial _{x}\left( \underset{n=0}{\overset{L+1}{\sum }}(\mathcal{W%
}_{(L)}^{nL}+B^{n})\frac{\partial \mathbf{H}_{L+1}}{\partial B^{n}}-\mathbf{H%
}_{L+1}\right) .
\end{equation*}%
The conservation law of the energy is%
\begin{equation*}
\partial _{t}\mathbf{H}_{L+1}=\partial _{x}\left[ \underset{k=0}{\overset{L+1%
}{\sum }}\underset{n=0}{\overset{L+1}{\sum }}\mathcal{W}_{(L)}^{kn}\frac{%
\partial \mathbf{H}_{L+1}}{\partial B^{k}}\frac{\partial \mathbf{H}_{L+1}}{%
\partial B^{n}}\right] .
\end{equation*}

By the construction all Hamiltonian structures for the Kupershmidt
hydrodynamic chain are written in the Liouville coordinates (see \textbf{%
\cite{Dubr+Nov}}, \textbf{\cite{Malt+Nov}}). Thus, all Hamiltonian
structures presented below can be easily verified by a substitution in above
formulas.

Let us re-compute the canonical Poisson bracket (nonzero components are
written below only; see \textbf{\cite{Dubr+Nov}})%
\begin{equation}
\{b^{i}(x),b^{i}(x^{\prime })\}=(\varepsilon _{i})^{-1}\delta ^{\prime
}(x-x^{\prime })  \label{cano}
\end{equation}%
via the moments $B_{(\gamma )}^{k}$ (see (\textbf{\ref{one}})). Then the
r.h.s. of the identity%
\begin{equation*}
\{B_{(\gamma )}^{k},B_{(\gamma )}^{n}\}=\beta ^{2}\underset{n=1}{\overset{N}{%
\sum }}\varepsilon _{i}[(b^{i})^{\beta (k+n+2\gamma )-2}\delta ^{\prime
}(x-x^{\prime })+(\beta (n+\gamma )-1)(b^{i})^{\beta (k+n+2\gamma
)-3}b_{x}^{i}\delta ^{\prime }(x-x^{\prime })]
\end{equation*}%
can be expressed via the moment $B_{(\gamma )}^{m}$ only, iff $\gamma
=2/\beta -L$, where $L$ is an integer. Thus, we have%
\begin{equation*}
\{B_{(2/\beta -L)}^{k},B_{(2/\beta -L)}^{n}\}=\beta ^{2}[(k-L+1/\beta
)B_{(2/\beta -L)}^{k+n-L}\partial _{x}+(n-L+1/\beta )\partial
_{x}B_{(2/\beta -L)}^{k+n-L}]\delta (x-x^{\prime }).
\end{equation*}%
Since the numeration of the moments $B_{(2/\beta -L)}^{k}$ starts from the
origin, then the above formula is valid if $k+n\geqslant L$. By this reason,
this family of the Poisson brackets will be slightly deformed on an extra
constant block (see (\textbf{\ref{y}})). Let us introduce the new notation $%
C_{(L+1)}^{k}=B_{(2/\beta -L)}^{k}/\beta $. Then the above main part ($%
k+n\geqslant L$) of the Poisson brackets presented below (see (\textbf{\ref%
{x}}))%
\begin{equation*}
\{C_{(L+1)}^{k},C_{(L+1)}^{n}\}=[(\beta (k-L)+1)C_{(L+1)}^{k+n-L}\partial
_{x}+(\beta (n-L)+1)\partial _{x}C_{(L+1)}^{k+n-L}]\delta (x-x^{\prime })
\end{equation*}%
is determined by the moments%
\begin{equation}
C_{(L+1)}^{k}=\frac{1}{\beta (k-L)+2}\underset{i=1}{\overset{M}{\sum }}%
\varepsilon _{i}(b^{i})^{\beta (k-L)+2}\text{, \ \ \ \ \ }k=0,1,2,...
\label{mom}
\end{equation}%
and equivalent the Kupershmidt Poisson bracket (\textbf{\ref{x}}) under the
re-numeration $C_{(L+1)}^{k}\rightarrow C_{(1)}^{k-L}$.

Let us consider $K+M$ component hydrodynamic type system (\textbf{\ref{t}})

\begin{eqnarray}
\partial _{t}C_{(K+1)}^{k} &=&\partial _{x}C_{(K+1)}^{k+1}+C^{0}\partial
_{x}C_{(K+1)}^{k}+[2+\beta (k-K)]C_{(K+1)}^{k}C_{x}^{0}\text{, \ \ \ }%
k=0,1,...,K-2,  \notag \\
&&  \notag \\
\partial _{t}C_{(K+1)}^{K-1} &=&\frac{1}{2}\partial _{x}\left( \sum
\varepsilon _{i}(b^{i})^{2}\right) +C^{0}\partial
_{x}C_{(K+1)}^{K-1}+(2-\beta )C_{(K+1)}^{K-1}C_{x}^{0}\text{,}  \label{j} \\
&&  \notag \\
b_{t}^{k} &=&\partial _{x}\left( \frac{(b^{k})^{\beta +1}}{\beta +1}%
+C^{0}b^{k}\right) \text{, \ \ \ \ \ \ \ }k=1,2,...,M  \notag
\end{eqnarray}%
connected with the equation of the Riemann surface (\textbf{\ref{r}})%
\begin{equation*}
\lambda =p^{\beta (K+1)-2}+(K+1-\frac{2}{\beta })\left[ \beta \underset{k=0}{%
\overset{K-1}{\sum }}C_{(K+1)}^{k}p^{\beta (K-k)-2}+\underset{m=1}{\overset{M%
}{\sum }}\varepsilon _{m}\left( \frac{b^{m}}{p}\right) ^{2}F\left( 1,\frac{2%
}{\beta },1+\frac{2}{\beta },\left( \frac{b^{m}}{p}\right) ^{\beta }\right) %
\right] .
\end{equation*}%
Since $\lambda $ satisfies the \textit{linear} equation (\textbf{\ref{8}}),
one can replace $\lambda $ on an arbitrary function $\tilde{\lambda}(\lambda
)$. Let us re-scale $\lambda ^{\beta (K+1)-2}\rightarrow \lambda ^{\beta }$.

The above hydrodynamic type system (\textbf{\ref{j}}) can be written in the
Hamiltonian form%
\begin{equation}
b_{t}^{i}=\frac{1}{\varepsilon _{i}(\beta +1)}\partial _{x}\frac{\partial 
\mathbf{h}_{K+1}}{\partial b^{i}}\text{, \ \ \ \ \ \ }h_{t}^{k}=\frac{1}{%
\beta +1}\partial _{x}\frac{\partial \mathbf{h}_{K+1}}{\partial h^{K-1-k}},
\label{red}
\end{equation}%
where the first $K$ coefficients $h^{k}(\mathbf{C})$ of the B\"{u}%
rmann--Lagrange series (see (\textbf{\ref{tri}}), (\textbf{\ref{rim}}) and (%
\textbf{\ref{kan}}))%
\begin{equation*}
1-p\lambda ^{-1/\beta }=\underset{n=0}{\overset{\infty }{\sum }}h^{k}\lambda
^{-k}\text{, \ \ \ \ \ }\lambda \rightarrow \infty
\end{equation*}%
can be obtained from the above equation of the Riemann surface (see \textbf{%
\cite{Lavr}}). Here $\mathbf{h}_{K}=\Sigma \varepsilon
_{i}(b^{i})^{2}/2+\Sigma h^{k}h^{K-1-k}/2$ is a momentum density, the
Hamiltonian density is $\mathbf{h}_{K+1}$.

\textbf{Theorem 2}: \textit{The Kupershmidt hydrodynamic chain }(\textbf{\ref%
{1}}) \textit{has infinitely many local Hamiltonian structures. The
hydrodynamic type system }(\textbf{\ref{red}}) \textit{is the Hamiltonian
hydrodynamic reduction of the Kupershmidt hydrodynamic chain written in the
Liouville coordinates }$C_{(K+1)}^{k}$\textit{\ of }$K$\textit{th local
Hamiltonian structure.}

\textbf{Proof} contains two steps:

\textbf{1}. It is necessary to prove that, indeed,%
\begin{equation}
\mathbf{h}_{K}=\frac{1}{2}\underset{i=0}{\overset{M}{\sum }}\varepsilon
_{i}(b^{i})^{2}+\frac{1}{2}\underset{i=0}{\overset{K}{\sum }}h^{k}h^{K-1-k}.
\label{imp}
\end{equation}%
The existence of the quadratic relationship between one conservation law
density $\mathbf{h}_{K}$ and the conservation law densities $b^{i},h^{k}$
means (see \textbf{\cite{Maks+Tsar}}) the existence of the local Hamiltonian
structure (\textbf{\ref{red}}). Then the Hamiltonian density $\mathbf{h}%
_{K+1}$ can be found in quadratures.

\textbf{2}. Introducing the moments $C_{(K+1)}^{k}$ via the moment
decomposition (\textbf{\ref{mom}}), the Hamiltonian hydrodynamic type system
(\textbf{\ref{j}}) transforms to the Kupershmidt hydrodynamic chain (\textbf{%
\ref{fuk}})%
\begin{equation}
\partial _{t}C_{(K+1)}^{k}=\partial _{x}C_{(K+1)}^{k+1}+C^{0}\partial
_{x}C_{(K+1)}^{k}+[\beta (k-K)+2]C_{(K+1)}^{k}C_{x}^{0}\text{, \ \ \ }%
k=0,1,2...;  \label{.}
\end{equation}%
simultaneously, the Hamiltonian structure of the hydrodynamic type system (%
\textbf{\ref{red}}) transforms to the $K$th local Hamiltonian structure of
the Kupershmidt hydrodynamic chain. To avoid a complexity of computations in
general case, we restrict our consideration on the second and third local
Hamiltonian structures, while the first local Hamiltonian structure was
established in \textbf{\cite{Kuper}}.

\textbf{The \textit{first} local Hamiltonian structure }(see (\textbf{\ref{x}%
}))%
\begin{equation}
\partial _{t}C_{(1)}^{k}=\frac{1}{\beta +1}[(\beta k+1)C_{(1)}^{k+n}\partial
_{x}+(\beta n+1)\partial _{x}C_{(1)}^{k+n}]\frac{\delta \mathbf{\bar{H}}_{1}%
}{\delta C_{(1)}^{n}},  \label{hama}
\end{equation}%
\textit{of the Kupershmidt hydrodynamic chain}%
\begin{equation*}
\partial _{t}C_{(1)}^{k}=\partial _{x}C_{(1)}^{k+1}+C^{0}\partial
_{x}C_{(1)}^{k}+(\beta k+2)C_{(1)}^{k}C_{x}^{0}\text{, \ \ \ }k=0,1,2,...
\end{equation*}%
\textit{is determined by the Hamiltonian} $\mathbf{\bar{H}}_{1}=\int
[C_{(1)}^{1}+(\beta +1)(C^{0})^{2}/2]dx$\textit{\ and by} \textit{the
momentum} $\mathbf{\bar{H}}_{0}=\int C^{0}dx$\textit{.}

Since the point transformation $C_{(1)}^{k}(\mathbf{B})$ is invertible (see (%
\textbf{\ref{inver}})), then the Poisson bracket $\{C_{(1)}^{k}$, $%
C_{(1)}^{n}\}$ can be expressed via the moments $B^{k}$ (see (\textbf{\ref%
{can}})). Thus, this local Poisson bracket $\{B^{k}$, $B^{n}\}_{1}$
determines the \textbf{first} local Hamiltonian structure for the
Kupershmidt hydrodynamic chain for an arbitrary index $\beta $.

Indeed, since the moment decomposition (\textbf{\ref{mom}})%
\begin{equation*}
C_{(1)}^{k}=\frac{1}{\beta k+2}\underset{i=1}{\overset{M}{\sum }}\varepsilon
_{i}(b^{i})^{\beta k+2}\text{, \ \ \ \ \ }k=0,1,2,...
\end{equation*}%
yields the momentum density ($K=0$)%
\begin{equation*}
\mathbf{h}_{0}=C^{0}=\frac{1}{2}\underset{m=1}{\overset{M}{\sum }}%
\varepsilon _{m}(b^{m})^{2},
\end{equation*}%
then one should just re-compute the canonical Poisson bracket (\textbf{\ref%
{cano}}) via the moments $C_{(1)}^{k}$ and check the Jacobi identity (%
\textbf{\ref{jac}}). Then the Hamiltonian density $\mathbf{h}_{1}$ of the
hydrodynamic reductions (\textbf{\ref{3}}) (see (\textbf{\ref{red}}), $K=0$)%
\begin{equation*}
b_{t}^{i}=\frac{1}{\varepsilon _{i}(\beta +1)}\partial _{x}\frac{\partial 
\mathbf{h}_{1}}{\partial b^{i}}
\end{equation*}%
becomes back the Hamiltonian density $\mathbf{H}_{1}$ of the first
Hamiltonian structure of the Kupershmidt hydrodynamic chain.

All higher commuting flows (see (\textbf{\ref{hama}}))%
\begin{equation}
\partial _{t^{m}}C_{(1)}^{k}=\frac{1}{\beta m+1}[(\beta
k+1)C_{(1)}^{k+n}\partial _{x}+(\beta n+1)\partial _{x}C_{(1)}^{k+n}]\frac{%
\delta \mathbf{\bar{H}}_{m}}{\delta C_{(1)}^{n}}\text{, \ \ \ \ }m=2,3,...
\label{q}
\end{equation}%
are determined by the higher Hamiltonians $\mathbf{\bar{H}}_{m}=\int \mathbf{%
H}_{m}(C_{(1)}^{0},C_{(1)}^{1},...,C_{(1)}^{m})dx$.

\textbf{The \textit{second} local Hamiltonian structure}%
\begin{eqnarray*}
C_{t}^{0} &=&\frac{1}{\beta +1}\partial _{x}\frac{\delta \mathbf{\bar{H}}_{2}%
}{\delta C^{0}}\text{,} \\
&& \\
\partial _{t}C_{(2)}^{k} &=&\frac{1}{\beta +1}[(\beta k+1-\beta
)C_{(2)}^{k+n-1}\partial _{x}+(\beta n+1-\beta )\partial _{x}C_{(2)}^{k+n-1}]%
\frac{\delta \mathbf{\bar{H}}_{2}}{\delta C_{(2)}^{n}}\text{, \ \ }%
k,n\geqslant 1
\end{eqnarray*}%
\textit{of the Kupershmidt hydrodynamic chain}%
\begin{equation*}
\partial _{t}C_{(2)}^{k}=\partial _{x}C_{(2)}^{k+1}+C^{0}\partial
_{x}C_{(2)}^{k}+(\beta (k-1)+2)C_{(2)}^{k}C_{x}^{0}\text{, \ \ \ }k=0\text{, 
}1\text{, }2\text{, ...}
\end{equation*}%
\textit{is determined by the Hamiltonian} $\mathbf{\bar{H}}_{2}=\int
[C_{(2)}^{2}+(\beta +1)C^{0}C_{(2)}^{1}+(3-\beta )(\beta +1)(C^{0})^{3}/6]dx$%
, \textit{where the momentum is} $\mathbf{\bar{H}}_{1}=\int
(C_{(2)}^{1}+(C^{0})^{2}/2]dx$, \textit{the Casimir (annihilator of
corresponding Poisson bracket) is} $\mathbf{\bar{H}}_{0}=\int C^{0}dx$.

Since the moment decomposition (\textbf{\ref{mom}})%
\begin{equation*}
C_{(2)}^{k+1}=\frac{1}{\beta k+2}\underset{i=1}{\overset{M}{\sum }}%
\varepsilon _{i}(b^{i})^{\beta k+2}\text{, \ \ \ \ \ }k=0,1,2,...
\end{equation*}%
yields the momentum density ($K=1$)%
\begin{equation*}
\mathbf{h}_{1}=C_{(2)}^{1}+\frac{1}{2}(C^{0})^{2}\equiv \frac{1}{2}\underset{%
m=1}{\overset{M}{\sum }}\varepsilon _{m}(b^{m})^{2}+\frac{1}{2}(h^{0})^{2},
\end{equation*}%
where $h^{0}=C^{0}$, then one should just re-compute the canonical Poisson
bracket (see (\textbf{\ref{cano}}); nonzero components are written below
only)%
\begin{equation*}
\{b^{i}(x),b^{i}(x^{\prime })\}=\delta ^{\prime }(x-x^{\prime })\text{, \ \
\ \ \ }\{h^{0}(x),h^{0}(x^{\prime })\}=\delta ^{\prime }(x-x^{\prime })
\end{equation*}%
via the moments $C_{(2)}^{k}$ and check the Jacobi identity (\textbf{\ref%
{jac}}). Then the Hamiltonian density $\mathbf{h}_{2}$ of the hydrodynamic
reductions (\textbf{\ref{red}})%
\begin{equation*}
b_{t}^{i}=\partial _{x}\left( \frac{(b^{i})^{\beta +1}}{\beta +1}%
+h^{0}b^{i}\right) \text{, \ \ \ \ \ \ \ \ }h_{t}^{0}=\partial _{x}\left( 
\frac{1}{2}\sum \varepsilon _{n}(b^{n})^{2}+\frac{3-\beta }{2}%
(h^{0})^{2}\right)
\end{equation*}%
written in the Hamiltonian form%
\begin{equation*}
b_{t}^{i}=\frac{1}{\varepsilon _{i}(\beta +1)}\partial _{x}\frac{\partial 
\mathbf{h}_{2}}{\partial b^{i}}\text{, \ \ \ \ \ \ }h_{t}^{0}=\frac{1}{\beta
+1}\partial _{x}\frac{\partial \mathbf{h}_{2}}{\partial h^{0}}
\end{equation*}%
becomes back the Hamiltonian density $\mathbf{H}_{2}$ of the second
Hamiltonian structure of the Kupershmidt hydrodynamic chain.

Since the point transformation $C_{(2)}^{k}(\mathbf{B})$ is invertible (see (%
\textbf{\ref{inver}})), then the Poisson bracket $\{C_{(2)}^{k}$, $%
C_{(2)}^{n}\}$ can be expressed via the moments $B^{k}$. Thus, this local
Poisson bracket $\{B^{k}$, $B^{n}\}_{2}$ determines the \textbf{second}
local Hamiltonian structure for the Kupershmidt hydrodynamic chain for an
arbitrary index $\beta $.

\textbf{The \textit{third} local Hamiltonian structure}.

\textit{The Kupershmidt hydrodynamic chain}%
\begin{equation*}
\partial _{t}C_{(3)}^{k}=\partial _{x}C_{(3)}^{k+1}+C^{0}\partial
_{x}C_{(3)}^{k}+(\beta (k-2)+2)C_{(3)}^{k}C_{x}^{0}\text{, \ \ \ }k=0\text{, 
}1\text{, }2\text{, ...}
\end{equation*}%
\textit{has the local Hamiltonian structure}%
\begin{eqnarray*}
C_{t}^{0} &=&\frac{1}{\beta +1}\partial _{x}\frac{\delta \mathbf{\bar{H}}_{3}%
}{\delta C_{(3)}^{1}}\text{,\ \ \ }\partial _{t}C_{(3)}^{1}=\frac{1}{\beta +1%
}\partial _{x}\frac{\delta \mathbf{\bar{H}}_{3}}{\delta C^{0}}+\frac{\beta -1%
}{\beta +1}(C^{0}\partial _{x}+\partial _{x}C^{0})\frac{\delta \mathbf{\bar{H%
}}_{3}}{\delta C_{(3)}^{1}}\text{,} \\
&& \\
\partial _{t}C_{(3)}^{k} &=&\frac{1}{\beta +1}[(\beta k+1-2\beta
)C_{(3)}^{k+n-2}\partial _{x}+(\beta n+1-2\beta )\partial
_{x}C_{(3)}^{k+n-2}]\frac{\delta \mathbf{\bar{H}}_{3}}{\delta C_{(3)}^{n}}%
\text{, \ \ }k,n\geqslant 2,
\end{eqnarray*}%
\textit{where the momentum is} $\mathbf{\bar{H}}_{2}=\int
[C_{(3)}^{2}+C^{0}C_{(3)}^{1}+(1-\beta )(C^{0})^{3}/2]dx$, \textit{two
Casimirs are }$\mathbf{\bar{H}}_{1}=\int [C_{(3)}^{1}+(1-\beta
)(C^{0})^{2}/2]dx$ \textit{and} $\mathbf{\bar{H}}_{0}=\int C^{0}dx$; \textit{%
the Hamiltonian is} $\mathbf{\bar{H}}_{3}=\int [C_{(3)}^{3}+(\beta
+1)C^{0}C_{(3)}^{2}+\frac{\beta +1}{2}(C_{(3)}^{1})^{2}+\frac{(3-2\beta
)(\beta +1)}{2}(C^{0})^{2}C_{(3)}^{1}+\frac{5(2\beta -3)(\beta ^{2}-1)}{24}%
(C^{0})^{4}]dx$.

Since the moment decomposition (\textbf{\ref{mom}})%
\begin{equation*}
C_{(2)}^{k+2}=\frac{1}{\beta k+2}\underset{i=1}{\overset{M}{\sum }}%
\varepsilon _{i}(b^{i})^{\beta k+2}\text{, \ \ \ \ \ }k=0,1,2,...
\end{equation*}%
yields the momentum density ($K=2$)%
\begin{equation*}
\mathbf{h}_{2}=C_{(3)}^{2}+C^{0}C_{(3)}^{1}+\frac{1-\beta }{2}%
(C^{0})^{3}\equiv \frac{1}{2}\underset{m=1}{\overset{M}{\sum }}\varepsilon
_{m}(b^{m})^{2}+h^{0}h^{1},
\end{equation*}%
where $h^{0}=C^{0}$, $h^{1}=C_{(3)}^{1}+(1-\beta )(C^{0})^{2}/2$, then one
should just re-compute the canonical Poisson bracket (see (\textbf{\ref{cano}%
}); nonzero components are written below only)%
\begin{equation*}
\{b^{i}(x),b^{i}(x^{\prime })\}=\delta ^{\prime }(x-x^{\prime })\text{, \ \
\ \ \ }\{h^{0}(x),h^{1}(x^{\prime })\}=\{h^{1}(x),h^{0}(x^{\prime
})\}=\delta ^{\prime }(x-x^{\prime })
\end{equation*}%
via the moments $C_{(2)}^{k}$ and check the Jacobi identity (\textbf{\ref%
{jac}}). Then the Hamiltonian density $\mathbf{h}_{3}$ of the hydrodynamic
reductions (\textbf{\ref{red}})%
\begin{eqnarray*}
b_{t}^{i} &=&\partial _{x}\left( \frac{(b^{i})^{\beta +1}}{\beta +1}%
+h^{0}b^{i}\right) \text{, \ \ \ }h_{t}^{0}=\partial _{x}[h^{1}+\frac{%
2-\beta }{2}(h^{0})^{2}]\text{,} \\
&& \\
h_{t}^{1} &=&\partial _{x}\left( \frac{1}{2}\sum \varepsilon
_{n}(b^{n})^{2}+(2-\beta )h^{0}h^{1}+\frac{\beta (\beta -1)}{6}%
(h^{0})^{3}\right)
\end{eqnarray*}%
written in the Hamiltonian form%
\begin{equation*}
b_{t}^{i}=\frac{1}{\varepsilon _{i}(\beta +1)}\partial _{x}\frac{\partial 
\mathbf{h}_{3}}{\partial b^{i}}\text{, \ \ \ \ \ \ }h_{t}^{0}=\frac{1}{\beta
+1}\partial _{x}\frac{\partial \mathbf{h}_{3}}{\partial h^{1}}\text{, \ \ \
\ \ \ }h_{t}^{1}=\frac{1}{\beta +1}\partial _{x}\frac{\partial \mathbf{h}_{3}%
}{\partial h^{0}}
\end{equation*}%
becomes back the Hamiltonian density $\mathbf{H}_{3}$ of the second
Hamiltonian structure of the Kupershmidt hydrodynamic chain.

Since the point transformation $C_{(3)}^{k}(\mathbf{B})$ is invertible (see (%
\textbf{\ref{inver}})), then the Poisson bracket $\{C_{(3)}^{k}$, $%
C_{(3)}^{n}\}$ can be expressed via the moments $B^{k}$. Thus, this local
Poisson bracket $\{B^{k}$, $B^{n}\}_{3}$ determines the \textbf{third} local
Hamiltonian structure for the Kupershmidt hydrodynamic chain for an
arbitrary index $\beta $.

In the same way all other local Hamiltonian structures can be constructed.
The Kupershmidt hydrodynamic chain (\textbf{\ref{.}}) is associated with the
Poisson $K-$bracket, where $K$ Casimirs $h^{k}$ are \textit{homogeneous
polynomials} with respect to the first moments $C_{(K+1)}^{k}$, $%
k=0,1,...,K-1$. All other higher moments are functions of the field
variables $b^{k}$ (see (\textbf{\ref{mom}}))%
\begin{equation}
C_{(K+1)}^{k+K}=\frac{1}{\beta k+2}\underset{i=1}{\overset{M}{\sum }}%
\varepsilon _{i}(b^{i})^{\beta k+2}\text{, \ \ \ \ \ }k=0,1,2,...
\label{dec}
\end{equation}%
Corresponding Poisson brackets $\{C_{(K+1)}^{k}$, $C_{(K+1)}^{n}\}$ under
the invertible point transformation (\textbf{\ref{inver}}) can be written
via the same set of the field variables $B^{k}$. These Poisson brackets $%
\{B^{k}$, $B^{n}\}_{K+1}$ determine the Kupershmidt hydrodynamic chain (%
\textbf{\ref{can}})%
\begin{equation*}
B_{t}^{k}=\frac{1}{\beta +1}\{B^{k}\text{, }\mathbf{\bar{H}}_{K+1}\}_{K+1}.
\end{equation*}%
by the Hamiltonians $\mathbf{\bar{H}}_{K+1}$, $K=0,1,2,...$, respectively.
All higher commuting flows also have the same set of the Hamiltonian
structures%
\begin{equation*}
B_{t^{m}}^{k}=\frac{1}{\beta m+1}\{B^{k}\text{, }\mathbf{\bar{H}}%
_{K+m}\}_{K+1}\text{, \ \ \ \ \ }m=0,1,2,...
\end{equation*}

\textit{All other Hamiltonian structures are \textbf{nonlocal}} (see \textbf{%
\cite{Fer+trans}}). For instance, the simplest such \textit{nonlocal}
Hamiltonian structure is associated with the Kupershmidt hydrodynamic chain
written in the form (cf. (\textbf{\ref{.}}))%
\begin{equation}
\partial _{t}C_{(0)}^{k}=\partial _{x}C_{(0)}^{k+1}+C^{0}\partial
_{x}C_{(0)}^{k}+(\beta (k+1)+2)C_{(0)}^{k}C_{x}^{0}\text{, \ \ \ }k=0\text{, 
}1\text{, }2\text{, ...}  \label{s}
\end{equation}

\textbf{Theorem 3}: \textit{The Kupershmidt hydrodynamic chain} (\textbf{\ref%
{s}}) \textit{has the nonlocal Hamiltonian structure}%
\begin{equation*}
\partial _{t}C_{(0)}^{k}=\{C_{(0)}^{k}\text{, }\mathbf{\bar{H}}_{0}\}\equiv 
\frac{1}{\beta +1}[(\beta k+\beta +1)C_{(0)}^{k+n+1}\partial _{x}+(\beta
n+\beta +1)\partial _{x}C_{(0)}^{k+n+1}
\end{equation*}%
\begin{equation*}
+\frac{1}{\beta }[(\beta k+\beta +2)(\beta n+\beta
+2)C_{(0)}^{k}C_{(0)}^{n}\partial _{x}+(\beta k+\beta +2)(\beta n+\beta
+1)C_{(0)}^{k}(C_{(0)}^{n})_{x}
\end{equation*}%
\begin{equation*}
+(\beta n+\beta +2)C_{(0)}^{n}(C_{(0)}^{k})_{x}-(C_{(0)}^{k})_{x}\partial
_{x}^{-1}(C_{(0)}^{n})_{x}]]\frac{\delta \mathbf{\bar{H}}_{0}}{\delta
C_{(0)}^{n}},
\end{equation*}%
\textit{where the Hamiltonian is} $\mathbf{\bar{H}}_{0}=\int C^{0}dx$.

\textbf{Proof}: The Kupershmidt hydrodynamic chain (\textbf{\ref{s}})
contains the hydrodynamic reduction (\textbf{\ref{3}})%
\begin{equation}
b_{t}^{i}=\partial _{x}\left( \frac{(b^{i})^{\beta +1}}{\beta +1}%
+C^{0}b^{i}\right) ,  \label{c}
\end{equation}%
where the moment decomposition is given by (cf. (\textbf{\ref{dec}}))%
\begin{equation*}
C_{(0)}^{k-1}=\frac{1}{\beta k+2}\underset{i=1}{\overset{M}{\sum }}%
\varepsilon _{i}(b^{i})^{\beta k+2}\text{, \ \ \ \ \ }k=1,2,...
\end{equation*}%
However, while the hydrodynamic reductions (\textbf{\ref{two}}) considered
in all above examples have $M$ independent field variables $b^{k}$, this
hydrodynamic type system has $M-1$ independent field variables only. Indeed,
the above hydrodynamic reduction (\textbf{\ref{c}}) has the \textit{quadratic%
} constraint%
\begin{equation*}
\underset{n=1}{\overset{M}{\sum }}\varepsilon _{n}(b^{n})^{2}=-1
\end{equation*}%
fixing the nonlocal Hamiltonian structure associated with a metric of
constant curvature if $\gamma =2/\beta +1$ (see details in \textbf{\cite%
{Fer+Mokh}}, \textbf{\cite{Maks+cc}}, \textbf{\cite{Maks+Tsar}}). Thus, the
above hydrodynamic type system can be written in the Hamiltonian form%
\begin{equation*}
b_{t}^{i}=\frac{1}{\beta +1}\partial _{x}\left[ \frac{1}{\varepsilon _{i}}%
\frac{\partial \mathbf{h}_{0}}{\partial b^{i}}+b^{i}\left( \underset{k=1}{%
\overset{M-1}{\sum }}b^{m}\frac{\partial \mathbf{h}_{0}}{\partial b^{m}}-%
\mathbf{h}_{0}\right) \right] \text{, \ \ \ }i=1,2,...,M-1\text{,}
\end{equation*}%
where $b^{M}(b^{1},b^{2},...,b^{M-1})$ can be expressed from the above
quadratic constraint.

A straightforward re-calculation yields the corresponding nonlocal
Hamiltonian structure of the Kupershmidt hydrodynamic chain (\textbf{\ref{s}}%
).

This is the first example of nonlocal Hamiltonian structures for
hydrodynamic chains. A direct verification of the Jacobi identity is not so
simple. However, under special reciprocal transformation this nonlocal
Hamiltonian structure becomes local (see \textbf{\cite{Maks+cc}}). Such
reciprocal transformations for nonlocal Hamiltonian structures will be
considered elsewhere.

\section{The extended Kupershmidt chains}

The Kupershmidt hydrodynamic chain (\textbf{\ref{fuk}}) has $N$ component
hydrodynamic reductions (\textbf{\ref{two}}) determined by the moment
decomposition (\textbf{\ref{one}}). This moment decomposition is defined for
any values of the index $k$. However, (\textbf{\ref{one}}) has been used for 
$k=0,1,2...$ only. Let us \textit{remove} this restriction.

\textbf{Definition 2}: \textit{The hydrodynamic chain}%
\begin{equation}
\partial _{t}B_{(\gamma )}^{k}=\partial _{x}B_{(\gamma )}^{k+1}+\frac{1}{%
\beta }B^{0}\partial _{x}B_{(\gamma )}^{k}+(k+\gamma )B_{(\gamma
)}^{k}B_{x}^{0}\text{, \ \ \ \ \ \ }k\in \mathbf{Z}  \label{ext}
\end{equation}%
\textit{is said to be the \textbf{extended} Kupershmidt hydrodynamic chain}.

(\textbf{\ref{ext}}) has the \textit{same} set of hydrodynamic reductions as
(\textbf{\ref{fuk}}). Thus, the extended Kupershmidt hydrodynamic chain is a
natural generalization of the Kupershmidt hydrodynamic chain.

As well as the hydrodynamic type system (\textbf{\ref{two}}), the extended
Kupershmidt hydrodynamic chain has \textit{negative} conservation laws and 
\textit{negative} commuting flows. For instance, the couple of the first
negative conservation laws is%
\begin{equation*}
\partial _{t^{1}}\mathbf{H}_{-1}=\partial _{t^{0}}\left( \frac{B^{0}}{\beta }%
\mathbf{H}_{-1}\right) \text{, \ \ \ }\partial _{t^{1}}\mathbf{H}%
_{-2}=\partial _{t^{0}}\left( \frac{\beta (\gamma -1)}{\beta +1}(\mathbf{H}%
_{-1})^{\beta +1}+\frac{B^{0}}{\beta }(\mathbf{H}_{-1})^{\beta (2-\gamma
)+1}B_{(\gamma )}^{-2}\right) ,
\end{equation*}%
where $t^{1}\equiv t$, $t^{0}\equiv x$ and 
\begin{equation*}
\mathbf{H}_{-1}=\left( 1+(\gamma -1)B_{(\gamma )}^{-1}\right) ^{\frac{1}{%
\beta (\gamma -1)}}\text{, \ \ \ \ \ }\mathbf{H}_{-2}=B_{(\gamma )}^{-2}(%
\mathbf{H}_{-1})^{\beta (2-\gamma )+1}.
\end{equation*}%
The extended Kupershmidt hydrodynamic chain (\textbf{\ref{ext}}) and its
commuting flows have the same set of local and nonlocal Hamiltonian
structures. For instance, the first \textit{negative} commuting flow%
\begin{equation}
\partial _{t^{-1}}B_{(\gamma )}^{k}=(\mathbf{H}_{-1})^{\beta -2}[\mathbf{H}%
_{-1}\partial _{x}B_{(\gamma )}^{k-1}-\beta (k+\gamma -1)B_{(\gamma
)}^{k-1}\partial _{x}\mathbf{H}_{-1}]  \label{neg}
\end{equation}%
can be determined by the first local Hamiltonian structure (\textbf{\ref{q}})%
\begin{equation*}
\partial _{t^{-1}}C_{(1)}^{k}=\frac{1}{1-\beta }[(\beta
k+1)C_{(1)}^{k-1}\partial _{x}+(1-\beta )\partial _{x}C_{(1)}^{k-1}]\frac{%
\delta \mathbf{\bar{H}}_{-1}}{\delta C_{(1)}^{-1}}\text{,\ \ \ \ \ }k\in 
\mathbf{Z}\mathcal{,}
\end{equation*}%
where the first \textit{negative} Hamiltonian is $\mathbf{\bar{H}}_{-1}=\int
\left( 1+(2/\beta -1)C_{(1)}^{-1}\right) ^{\frac{1}{2-\beta }}dx$.

\textbf{Remark}: Let us restrict our consideration of the above hydrodynamic
chain (\textbf{\ref{neg}}) on \textit{purely negative} values of the
discrete variable $k$. The corresponding hydrodynamic chain is%
\begin{equation*}
\partial _{y}\tilde{B}_{(\gamma )}^{k}=(\tilde{B}_{(\gamma
)}^{0}+\varepsilon )^{\frac{\beta -1}{\beta (\gamma -1)}-1}\left[ (\tilde{B}%
_{(\gamma )}^{0}+\varepsilon )\partial _{x}\tilde{B}_{(\gamma )}^{k+1}+\frac{%
k+2-\gamma }{\gamma -1}\tilde{B}_{(\gamma )}^{k+1}\partial _{x}\tilde{B}%
_{(\gamma )}^{0}\right] \text{, \ \ \ \ \ }k=0,1,2,...,
\end{equation*}%
where $y=t^{-1}\varepsilon ^{\frac{1-\beta }{\beta (\gamma -1)}}$, $\tilde{B}%
_{(\gamma )}^{k}=\varepsilon (\gamma -1)B_{(\gamma )}^{-k-1}$ and $%
\varepsilon $ is an arbitrary constant removable by the shift $\tilde{B}%
_{(\gamma )}^{0}+\varepsilon \rightarrow \tilde{B}_{(\gamma )}^{0}$. This
hydrodynamic chain was studied in \textbf{\cite{Blaszak}}, \textbf{\cite%
{Kuper1}}, \textbf{\cite{Manasa}}, \textbf{\cite{Maks+algebr}} and \textbf{%
\cite{Maks+Hamch}}. This constant $\varepsilon $ plays an important role in
a non-degeneracy of the corresponding hydrodynamic reductions (\textbf{\ref%
{one}}). If $\varepsilon =0$, then the extra constraint $\Sigma \varepsilon
_{k}(b^{k})^{\beta \gamma }=\limfunc{const}$ appears.

The simplest $N$ parametric family of $N$ component hydrodynamic reductions
of the hydrodynamic chain (\textbf{\ref{neg}})%
\begin{equation}
b_{t^{-1}}^{i}=\partial _{x}\left[ \frac{(\mathbf{H}_{-1}/b^{i})^{\beta -1}}{%
1-\beta }\right]  \label{negat}
\end{equation}%
commutes with (\textbf{\ref{two}}). The compatibility condition $\partial
_{t}(\partial _{t^{-1}}p)=\partial _{t^{-1}}(\partial _{t}p)$ of the
generating functions of conservation laws (\textbf{\ref{tri}}) and (we
replace $b^{i}\rightarrow p$, see \textbf{\cite{Maks+algebr}})%
\begin{equation}
p_{t^{-1}}=\partial _{x}\left[ \frac{(p/\mathbf{H}_{-1})^{1-\beta }}{1-\beta 
}\right]  \label{cona}
\end{equation}%
yields (see \textbf{\cite{Manasa}}) 2+1 quasilinear system%
\begin{equation*}
\partial _{t^{-1}}\mathbf{H}_{0}=\partial _{x}\frac{(\mathbf{H}_{-1})^{\beta
-1}}{\beta -1}\text{, \ \ \ \ \ \ }\partial _{t}\mathbf{H}_{-1}=\partial
_{x}(\mathbf{H}_{0}\mathbf{H}_{-1}),
\end{equation*}%
where $\mathbf{H}_{0}=B^{0}/\beta $.

As well as in the second section a consistency of the generating function of
conservation laws (\textbf{\ref{cona}}) with the hydrodynamic type system (%
\textbf{\ref{negat}}) yields (\textbf{\ref{comp}}). Taking into account the
moment decomposition for the negative moments (see (\textbf{\ref{one}}))%
\begin{equation*}
B_{(\gamma )}^{-k}=\frac{1}{\gamma -k}\underset{i=1}{\overset{N}{\sum }}%
\varepsilon _{i}(b^{i})^{\beta (\gamma -k)}\text{, \ \ \ \ \ }\gamma \neq
0,1,2,...,
\end{equation*}%
the \textbf{another} equation of the Riemann mapping (cf. (\textbf{\ref{rim}}%
) and (\textbf{\ref{dif}}))%
\begin{equation}
\lambda =q^{1-\gamma }-(1-\gamma )\underset{k=1}{\overset{\infty }{\sum }}%
B_{(\gamma )}^{-k}q^{k-\gamma }\text{, \ \ \ \ \ \ }\lambda \rightarrow 0
\label{net}
\end{equation}%
can be found (the above moment decomposition degenerates according to (%
\textbf{\ref{exep}}) in the exceptional cases $\gamma =0,1,2,...$). As well
as in the \textit{positive} case (\textbf{\ref{inver}}) the negative moments 
$B_{(\gamma )}^{-k}$ with the distinct parameters $\gamma $ can be expressed
via each other by the invertible transformations%
\begin{equation*}
B_{(\gamma )}^{-1}=\frac{1-(1-B^{-1})^{1-\gamma }}{1-\gamma }\text{,\ \ \ \
\ \ \ \ }B_{(\gamma )}^{-2}=B^{-2}(1-B^{-1})^{-\gamma },...
\end{equation*}

If $\gamma =1$, the extended Kupershmidt hydrodynamic chain (\textbf{\ref%
{ext}})%
\begin{equation*}
\partial _{t}B_{(1)}^{k}=\partial _{x}B_{(1)}^{k+1}+\frac{1}{\beta }%
B^{0}\partial _{x}B_{(1)}^{k}+(k+1)B_{(1)}^{k}B_{x}^{0}\text{, \ \ \ \ \ \ }%
k\in \mathbf{Z}
\end{equation*}%
has the first negative conservation law%
\begin{equation*}
\partial _{t}\mathbf{H}_{-1}=\frac{1}{\beta }\partial _{x}\left( B^{0}%
\mathbf{H}_{-1}\right) ,
\end{equation*}%
where $\mathbf{H}_{-1}=e^{B_{(1)}^{-1}/\beta }$, and the first negative
commuting flow%
\begin{equation*}
\partial _{t^{-1}}B_{(1)}^{k}=e^{\frac{\beta -1}{\beta }B_{(1)}^{-1}}[%
\partial _{x}B_{(1)}^{k-1}-kB_{(1)}^{k-1}\partial _{x}B_{(1)}^{-1}].
\end{equation*}%
The negative moments $B_{(1)}^{-k}$ are connected with the equation of the
Riemann mapping (\textbf{\ref{net}}) reduced to (cf. (\textbf{\ref{log}}))%
\begin{equation*}
\lambda =\ln q-\underset{k=1}{\overset{\infty }{\sum }}B_{(1)}^{-k}q^{k-1}%
\text{ \ \ \ \ }\Leftrightarrow \text{ \ \ \ \ }\lambda =\frac{1}{q}\exp
\left( \underset{k=1}{\overset{\infty }{\sum }}B_{(1)}^{-k}q^{k-1}\right) 
\text{, \ \ \ \ }\lambda \rightarrow \infty \text{, \ \ }q\rightarrow 0.
\end{equation*}%
Since $B_{(1)}^{-1}=\beta \Sigma \varepsilon _{k}\ln b^{k}$ and $\Sigma
\varepsilon _{k}=0$ in this exceptional case ($\gamma =1$, see (\textbf{\ref%
{exep}})), then $\mathbf{H}_{-1}=\Pi (b^{k})^{\varepsilon _{k}}$.

\textbf{Summary}: The Kupershmidt hydrodynamic chain (\textbf{\ref{fuk}})
can be extended on the negative moments $B_{(\gamma )}^{-k}$. The extended
Kupershmidt hydrodynamic chain (\textbf{\ref{ext}}) has infinitely many
higher (positive) and lower (negative) conservation laws and commuting
flows. Two different asymptotic Riemann mappings (\textbf{\ref{rim}}) and (%
\textbf{\ref{net}}) are connected with positive and negative parts of these
hydrodynamic chains, respectively.

Let us associate the ``time'' indexes $m$ of commuting flows $\partial
_{t^{m}}B_{(\gamma )}^{k}$ and the moment indexes $k$ with the corresponding
dots ($m,k$) on the plane. By this reason the hierarchy of the Kupershmidt
hydrodynamic chain (including all conservation laws enumerated by the index $%
k$ and all commuting flows enumerated by the index $m$) we call the
Kupershmidt hydrodynamic \textit{lattice}.

\section{Linear transformation of independent variables}

In this section we describe the simplest discrete symmetry of the
Kupershmidt hydrodynamic lattice (see the corresponding transformation of
hydrodynamic reductions in \textbf{\cite{Maks+algebr}}). By this reason, the
generating function of conservation law densities $p^{(\beta )}$, all
conservation law densities $\mathbf{H}_{k}^{(\beta )}$ and all moments $%
B_{(\beta ,\gamma )}^{k}$ we write with the extra index $\beta $.

\textbf{Theorem 4}: \textit{The generating function of conservation laws} (%
\textbf{\ref{cona}})%
\begin{equation*}
p_{t^{-1}}^{(\beta )}=\partial _{t^{0}}\left[ \frac{\left( p^{(\beta )}/%
\mathbf{H}_{-1}^{(\beta )}\right) ^{1-\beta }}{1-\beta }\right]
\end{equation*}%
\textit{is invariant under the transformation} $t^{0}\leftrightarrow t^{-1}$%
. \textit{Then}%
\begin{equation*}
p^{(\tilde{\beta})}=\left( p^{(\beta )}/\mathbf{H}_{-1}^{(\beta )}\right)
^{1-\beta }\text{,\ \ \ \ \ \ \ }\mathbf{H}_{-1}^{(\tilde{\beta})}=\left( 
\mathbf{H}_{-1}^{(\beta )}\right) ^{\beta -1}\text{, \ \ \ \ \ \ }\tilde{B}%
_{(\tilde{\beta},\tilde{\gamma})}^{k}=B_{(\beta ,\gamma )}^{-k-2}\left( 
\mathbf{H}_{-1}^{(\beta )}\right) ^{\beta (k+2-\gamma )}\text{,}
\end{equation*}%
\textit{where}%
\begin{equation*}
\tilde{\beta}=\frac{\beta }{\beta -1}\text{, \ \ \ \ \ }\tilde{\gamma}%
=2-\gamma
\end{equation*}%
\textit{and} $\tilde{\varepsilon}_{i}=-\varepsilon _{i}$ \textit{in the
corresponding moment decompositions} (\textbf{\ref{one}}).

\textbf{Proof}: can be obtained by a straightforward substitution in (%
\textbf{\ref{neg}}) and (\textbf{\ref{cona}}).

\textbf{Corollary}: This transformation connects (\textbf{\ref{rim}}) with (%
\textbf{\ref{net}}). These equations of the Riemann mapping are related by%
\begin{equation*}
q^{(\tilde{\beta})}=\frac{\left( \mathbf{H}_{-1}^{(\beta )}\right) ^{\beta }%
}{q^{(\beta )}}.
\end{equation*}

\textbf{Remark}: If $\gamma =1$, then $\tilde{\gamma}=1$ and $\tilde{B}_{(%
\tilde{\beta},1)}^{k}=B_{(\beta ,1)}^{-k-2}\exp [(k+1)B_{(\beta ,1)}^{-1}]$.

Below we extend this discrete transformation $t^{0}\leftrightarrow t^{-1}$
on all other ``times'' $t^{k-1}\leftrightarrow t^{-k}$. Following the recipe
given in \textbf{\cite{Maks+vech}} (see also \textbf{\cite{Maks+algebr}})
one can seek a \textit{generating function of conservation laws and
commuting flows} in the form%
\begin{equation}
\partial _{\tau (\zeta )}p(\lambda )=\partial _{t^{0}}F(p(\lambda ),p(\zeta
)),  \label{total}
\end{equation}%
where $\partial _{\tau (\zeta )}$ is the so-called ``vertex'' operator (see,
for instance, \textbf{\cite{Bogdan}} and details below). The compatibility
conditions $\partial _{t^{1}}(\partial _{\tau (\zeta )}p(\lambda ))=\partial
_{\tau (\zeta )}(\partial _{t^{1}}p(\lambda ))$ and $\partial
_{t^{-1}}(\partial _{\tau (\zeta )}p(\lambda ))=\partial _{\tau (\zeta
)}(\partial _{t^{-1}}p(\lambda ))$ (see (\textbf{\ref{tri}}) and (\textbf{%
\ref{cona}})) yield%
\begin{equation}
\partial _{\tau (\zeta )}\mathbf{H}_{0}=\partial _{t^{0}}\frac{p^{\beta
-1}(\zeta )}{\beta -1}\text{, \ \ \ \ \ }\partial _{\tau (\zeta )}\mathbf{H}%
_{-1}=-\partial _{t^{0}}\frac{\mathbf{H}_{-1}}{p(\zeta )}\text{,}  \label{u}
\end{equation}%
where%
\begin{equation*}
dF(w)=\frac{dw}{w^{\beta }-1}\text{, \ \ \ \ \ }w=p(\lambda )/p(\zeta ).
\end{equation*}

Taking into account two distinct Riemann mappings (\textbf{\ref{rim}}) and (%
\textbf{\ref{net}}) one can introduce the inverse asymptotics $p(\zeta )$
and consistent formal series for the vertex operator $\partial _{\tau (\zeta
)}$%
\begin{eqnarray}
\partial _{\tau (\zeta )} &=&-\zeta ^{-1/\beta }\underset{k=0}{\overset{%
\infty }{\sum }}\zeta ^{-k}\partial _{t^{k}}\text{, \ \ \ \ \ \ \ \ \ }%
p(\zeta )=\zeta ^{1/\beta }\left( 1-\underset{k=1}{\overset{\infty }{\sum }}%
\zeta ^{-k}\mathbf{H}_{k-1}\right) \text{,}  \notag \\
&&  \label{ful} \\
\partial _{\tau (\zeta )} &=&\zeta ^{1/\beta }\underset{k=1}{\overset{\infty 
}{\sum }}\zeta ^{-k}\partial _{t^{-k}}\text{, \ \ \ \ \ \ \ \ \ \ \ }p(\zeta
)=\zeta ^{-1/\beta }\underset{k=0}{\overset{\infty }{\sum }}\zeta ^{-k}%
\mathbf{H}_{-k-1},  \notag
\end{eqnarray}%
when $\zeta \rightarrow \infty $.

Applying above series to \textit{the generating function of conservation
laws and commuting flows} (see (\textbf{\ref{total}})) one can obtain
infinite series of separate generating functions of conservation laws and
commuting flows (cf. (\textbf{\ref{tri}}), (\textbf{\ref{cona}}), (\textbf{%
\ref{u}}))%
\begin{eqnarray*}
\partial _{t^{1}}p &=&\partial _{t^{0}}\left( \frac{p^{\beta +1}}{\beta +1}+%
\mathbf{H}_{0}p\right) \text{, \ \ \ }\partial _{t^{2}}p=\partial
_{t^{0}}\left( \frac{p^{2\beta +1}}{2\beta +1}+\mathbf{H}_{0}p^{\beta +1}+(%
\mathbf{H}_{1}+(\mathbf{H}_{0})^{2})p\right) ,... \\
&& \\
\partial _{t^{-1}}p &=&\partial _{t^{0}}\left( \frac{(p/\mathbf{H}%
_{-1})^{1-\beta }}{1-\beta }\right) \text{, \ \ \ \ \ }\partial
_{t^{-2}}p=\partial _{t^{0}}\left( \frac{(p/\mathbf{H}_{-1})^{1-2\beta }}{%
1-2\beta }-\frac{\mathbf{H}_{-3}}{\mathbf{H}_{-2}}(p/\mathbf{H}%
_{-1})^{1-\beta }\right) ,... \\
&& \\
\partial _{\tau }\mathbf{H}_{1} &=&\partial _{t^{0}}\left( \frac{p^{2\beta
-1}}{2\beta -1}+\mathbf{H}_{0}p^{\beta -1}\right) \text{, \ \ \ }\partial
_{\tau }\mathbf{H}_{2}=\partial _{t^{0}}\left( \frac{p^{3\beta -1}}{3\beta -1%
}+\mathbf{H}_{0}p^{2\beta -1}+(\mathbf{H}_{1}+\frac{\beta }{2}(\mathbf{H}%
_{0})^{2})p^{\beta -1}\right) ,... \\
&& \\
\partial _{\tau }\mathbf{H}_{0} &=&\partial _{t^{0}}\frac{p^{\beta -1}}{%
\beta -1}\text{, \ \ \ \ }\partial _{\tau }\mathbf{H}_{-1}=-\partial _{t^{0}}%
\frac{\mathbf{H}_{-1}}{p}\text{,\ \ \ \ \ }\partial _{\tau }\mathbf{H}%
_{-2}=-\partial _{t^{0}}\left( \frac{\mathbf{H}_{-2}}{p}+\frac{(\mathbf{H}%
_{-1}/p)^{\beta +1}}{\beta +1}\right) ,...
\end{eqnarray*}%
Let us rewrite the generating function of conservation laws of the
Kupershmidt hydrodynamic chain together with the second line of above
formulas in the potential form%
\begin{equation*}
d\xi =...+\left( \frac{p^{\beta +1}}{\beta +1}+\mathbf{H}_{0}p\right)
dt^{1}+pdt^{0}+\frac{(p/\mathbf{H}_{-1})^{1-\beta }}{1-\beta }dt^{-1}+\left( 
\frac{(p/\mathbf{H}_{-1})^{1-2\beta }}{1-2\beta }-\frac{\mathbf{H}_{-3}}{%
\mathbf{H}_{-2}}(p/\mathbf{H}_{-1})^{1-\beta }\right) dt^{-2}+...
\end{equation*}%
The compatibility conditions $(\xi _{t^{-1}})_{t^{0}}=(\xi _{t^{0}})_{t^{-1}}
$ and $(\xi _{t^{-1}})_{t^{-2}}=(\xi _{t^{-2}})_{t^{-1}}$ yield the
generating functions of conservation laws (cf. (\textbf{\ref{tri}}) and (%
\textbf{\ref{cona}}))%
\begin{equation*}
\partial _{t^{0}}p^{(\tilde{\beta})}=\partial _{t^{-1}}\left[ \frac{\left(
p^{(\tilde{\beta})}/\mathbf{H}_{-1}^{(\tilde{\beta})}\right) ^{1-\tilde{\beta%
}}}{1-\tilde{\beta}}\right] \text{, \ \ \ \ \ \ \ }\partial _{t^{-2}}p^{(%
\tilde{\beta})}=\partial _{t^{-1}}\left[ \frac{(p^{(\tilde{\beta})})^{\tilde{%
\beta}+1}}{\tilde{\beta}+1}+\mathbf{H}_{0}^{(\tilde{\beta})}p^{(\tilde{\beta}%
)}\right] ,
\end{equation*}%
where (see the above theorem)%
\begin{equation*}
\tilde{\beta}=\frac{\beta }{\beta -1}\text{, \ \ \ \ \ \ \ }p^{(\tilde{\beta}%
)}=\left( \frac{p^{(\beta )}}{\mathbf{H}_{-1}^{(\beta )}}\right) ^{1-\beta }%
\text{, \ \ \ \ \ \ \ }\mathbf{H}_{-1}^{(\tilde{\beta})}=\left( \mathbf{H}%
_{-1}^{(\beta )}\right) ^{\beta -1}\text{,\ \ \ \ \ \ \ \ }\mathbf{H}_{0}^{(%
\tilde{\beta})}=(\beta -1)\frac{\mathbf{H}_{-2}^{(\beta )}}{\mathbf{H}%
_{-1}^{(\beta )}}\text{.}
\end{equation*}%
The substitution of the third and the fourth series (\textbf{\ref{ful}}) in
the second above equation%
\begin{equation*}
p^{(\tilde{\beta})}=\left( \frac{p^{(\beta )}}{\mathbf{H}_{-1}^{(\beta )}}%
\right) ^{1-\beta }
\end{equation*}%
yields explicit expressions $\mathbf{H}_{k}^{(\tilde{\beta})}$ via $\mathbf{H%
}_{n}^{(\beta )}$ (this is an invertible transformation) for any indexes $k$%
. For instance,%
\begin{eqnarray*}
\mathbf{\tilde{H}}_{1} &=&(\beta -1)\left( \frac{\mathbf{H}_{-3}}{\mathbf{H}%
_{-1}}-\frac{\beta }{2}\frac{(\mathbf{H}_{-2})^{2}}{(\mathbf{H}_{-1})^{2}}%
\right) \text{, \ \ }\mathbf{\tilde{H}}_{2}=(\beta -1)\left( \frac{\mathbf{H}%
_{-4}}{\mathbf{H}_{-1}}-\beta \frac{\mathbf{H}_{-2}\mathbf{H}_{-3}}{(\mathbf{%
H}_{-1})^{2}}+\frac{\beta (\beta +1)}{6}\frac{(\mathbf{H}_{-2})^{3}}{(%
\mathbf{H}_{-1})^{3}}\right) ,..., \\
&& \\
\mathbf{\tilde{H}}_{-2} &=&(\beta -1)\mathbf{H}_{0}\left( \mathbf{H}%
_{-1}\right) ^{\beta -1}\text{, \ \ \ \ \ \ \ \ \ }\mathbf{\tilde{H}}%
_{-3}=(\beta -1)\left( \mathbf{H}_{-1}\right) ^{\beta -1}\left( \mathbf{H}%
_{1}+\frac{\beta (\beta +1)}{2}\left( \mathbf{H}_{0}\right) ^{2}\right) ,...,
\end{eqnarray*}%
where for simplicity we use the temporary notation $\mathbf{H}_{k}=\mathbf{H}%
_{k}^{(\beta )}$, $\mathbf{\tilde{H}}_{k}=\mathbf{H}_{k}^{(\tilde{\beta})}$.

Thus, the Kupershmidt hydrodynamic \textit{lattice} (i.e. a whole family of
commuting hydrodynamic \textit{chains}) admits the transformation $\beta
\leftrightarrow \tilde{\beta}$, $t^{k}\leftrightarrow \tilde{t}^{-1-k}$, $%
k=0,\pm 1,\pm 2$, ... It means that the Kupershmidt hydrodynamic chain with
the index $\beta $ has infinitely many \textit{negative} conservation laws,
which are \textit{positive} conservation laws for the Kupershmidt
hydrodynamic chain with the index $\tilde{\beta}$. For instance, the
modified Benney chain ($\beta =1$) is \textit{matched together} with the
dispersionless limit of the discrete KP hierarchy ($\beta =\infty $). The
case $\beta =2$ (dispersionless limit of the Veselov--Novikov equation) is 
\textit{invariant} under the above transformation.

\section{Reciprocal transformations}

In the theory of integrable dispersive and dispersionless systems the
concept of \textit{reciprocal transformations} was introduced by S.A.
Chaplygin (see details in \textbf{\cite{Rogers}} and \textbf{\cite{Yanenko}}%
). Also reciprocal transformations are useful in an application to
hydrodynamic chains (see first examples in \textbf{\cite{Fer+Dav}} and 
\textbf{\cite{Maks+Egor}}). Let us substitute the Taylor series (\textbf{\ref%
{ful}}) in the generating function of commuting flows%
\begin{equation*}
\partial _{\tau (\zeta )}\mathbf{H}_{-1}=-\partial _{t^{0}}\frac{\mathbf{H}%
_{-1}}{p(\zeta )}
\end{equation*}%
and rewrite corresponding conservation laws in the potential form%
\begin{equation*}
dy^{0}=...+[\frac{\mathbf{H}_{-3}}{\mathbf{H}_{-1}}-\frac{(\mathbf{H}%
_{-2})^{2}}{(\mathbf{H}_{-1})^{2}}]dt^{-2}+\frac{\mathbf{H}_{-2}}{\mathbf{H}%
_{-1}}dt^{-1}+\mathbf{H}_{-1}[dt^{0}+\mathbf{H}_{0}dt^{1}+(\mathbf{H}_{1}+(%
\mathbf{H}_{0})^{2})dt^{2}+...].
\end{equation*}%
The reciprocal transformation $dy^{0}=...$ and $dy^{k}=dt^{-k},k=\pm 1,\pm
2,...$ applied to the generating functions of conservation laws (see the
previous section) yields the generating functions of conservation laws%
\begin{eqnarray*}
\partial _{y^{1}}\tilde{p} &=&\partial _{y^{0}}\left( \frac{\tilde{p}%
^{1-\beta }}{1-\beta }+\mathbf{\tilde{H}}_{0}\tilde{p}\right) \text{, \ \ \ }%
\partial _{y^{2}}\tilde{p}=\partial _{y^{0}}\left( \frac{\tilde{p}^{1-2\beta
}}{1-2\beta }+\mathbf{\tilde{H}}_{0}\tilde{p}^{1-\beta }+(\mathbf{\tilde{H}}%
_{1}+(\mathbf{\tilde{H}}_{0})^{2})\tilde{p}\right) \text{, ...} \\
&& \\
\partial _{y^{-1}}\tilde{p} &=&\partial _{y^{0}}\left( \frac{(\tilde{p}/%
\mathbf{\tilde{H}}_{-1})^{\beta +1}}{\beta +1}\right) \text{, \ \ \ \ \ }%
\partial _{y^{-2}}\tilde{p}=\partial _{y^{0}}\left( \frac{(\tilde{p}/\mathbf{%
\tilde{H}}_{-1})^{2\beta +1}}{2\beta +1}-\frac{\mathbf{\tilde{H}}_{-2}}{%
\mathbf{\tilde{H}}_{-1}}(\tilde{p}/\mathbf{\tilde{H}}_{-1})^{\beta
+1}\right) \text{, ...}
\end{eqnarray*}%
where%
\begin{equation*}
\tilde{p}=\frac{p}{\mathbf{H}_{-1}}\text{, \ }\mathbf{\tilde{H}}_{-1}=\frac{1%
}{\mathbf{H}_{-1}}\text{,\ \ }\mathbf{\tilde{H}}_{-2}=-\frac{\mathbf{H}_{0}}{%
\mathbf{H}_{-1}}\text{,\ \ }\mathbf{\tilde{H}}_{-3}=-\frac{\mathbf{H}_{1}}{%
\mathbf{H}_{-1}},...,\text{ \ \ }\mathbf{\tilde{H}}_{0}=-\frac{\mathbf{H}%
_{-2}}{\mathbf{H}_{-1}}\text{,\ \ }\mathbf{\tilde{H}}_{1}=-\frac{\mathbf{H}%
_{-3}}{\mathbf{H}_{-1}},...
\end{equation*}

\textbf{Theorem 5}: \textit{The Kupershmidt hydrodynamic chains} (\textbf{%
\ref{fuk}})%
\begin{eqnarray*}
\partial _{t^{1}}B_{(\beta ,\gamma )}^{n} &=&\partial _{t^{0}}B_{(\beta
,\gamma )}^{n+1}+\frac{1}{\beta }B^{0}\partial _{t^{0}}B_{(\beta ,\gamma
)}^{n}+(n+\gamma )B_{(\beta ,\gamma )}^{n}\partial _{t^{0}}B^{0}\text{, \ \
\ }n=0,1,2,..., \\
&& \\
\partial _{y^{-1}}\tilde{B}_{(\tilde{\beta},\tilde{\gamma})}^{k} &=&(\mathbf{%
\tilde{H}}_{-1})^{\tilde{\beta}-2}[\mathbf{\tilde{H}}_{-1}\partial _{y^{0}}%
\tilde{B}_{(\tilde{\beta},\tilde{\gamma})}^{k-1}-\tilde{\beta}(k+\tilde{%
\gamma}-1)\tilde{B}_{(\tilde{\beta},\tilde{\gamma})}^{k-1}\partial _{y^{0}}%
\mathbf{\tilde{H}}_{-1}]\text{, \ \ \ }n=0,1,2,...
\end{eqnarray*}%
\textit{are related by the above reciprocal transformation, where (cf. the
theorem from the previous section)}%
\begin{equation*}
\tilde{B}_{(\tilde{\beta},\tilde{\gamma})}^{k}=B_{(\beta ,\gamma
)}^{-k-2}\left( \mathbf{H}_{-1}^{(\beta )}\right) ^{\beta (k+2-\gamma )}%
\text{, \ \ \ \ \ \ }\tilde{\beta}=-\beta \text{, \ \ \ \ \ }\tilde{\gamma}%
=2-\gamma
\end{equation*}%
\textit{and} $\tilde{\varepsilon}_{i}=-\varepsilon _{i}$ \textit{in the
corresponding moment decompositions} (\textbf{\ref{one}}).

\textbf{Proof}: can be obtained by a straightforward substitution of the
above three formulas in the above hydrodynamic chains.

\textbf{Remark}: The case $\beta =1/N$ was considered in \textbf{\cite%
{Blaszak}} (see also \textbf{\cite{Manasa}}). The Kupershmidt hydrodynamic
chain (the first above) and its \textbf{higher} commuting flows in such case
are called ``$N-$dmKP hierarchy''. The Kupershmidt hydrodynamic chain (the
second above) and its \textbf{lower} commuting flows are called ``$N-$dDym
hierarchy''. We would like to emphasize again here that corresponding
hydrodynamic chains are members of a sole hierarchy. For the case $\beta
=1/N $ all together these commuting flows are called ``$N-$dToda hierarchy''
(see \textbf{\cite{Manasa}}).

Let us take into account the \textit{discrete transformation of independent
variables} described in the previous section. Then the index $\beta $ can be
changed in a combination of these both transformations%
\begin{eqnarray*}
\beta _{(0)} &\rightarrow &\beta _{(1)}=-\beta _{(0)}\text{, \ \ }\beta
_{(1)}\rightarrow \beta _{(2)}=\frac{\beta _{(1)}}{\beta _{(1)}-1}\text{, \
\ \ }\beta _{(2)}\rightarrow \beta _{(3)}=-\beta _{(2)},... \\
&& \\
\beta _{(0)} &\rightarrow &\beta _{(-1)}=\frac{\beta _{(0)}}{\beta _{(0)}-1}%
\text{, \ \ \ }\beta _{(-1)}\rightarrow \beta _{(-2)}=-\beta _{(-1)}\text{,
\ \ \ }\beta _{(-2)}\rightarrow \beta _{(-3)}=\frac{\beta _{(-2)}}{\beta
_{(-2)}-1},...
\end{eqnarray*}%
Thus,%
\begin{equation*}
\beta _{(2K)}=-\beta _{(2K+1)}=\frac{\beta _{(0)}}{\beta _{(0)}K+1}\text{, \
\ \ \ \ }K=0,\pm 1,\pm 2,...
\end{equation*}%
Let us start, for instance, from the dispersionless limit of the discrete
DKP hierarchy, where the \textit{initial} hydrodynamic chain is (see, for
instance, \textbf{\cite{Gib+Yu}})%
\begin{equation*}
B_{t}^{n}=B_{x}^{n+1}+nB^{n}B_{x}^{0}\text{, \ \ \ }n=0,...
\end{equation*}%
It means that $\beta _{(0)}=\infty $. Since%
\begin{equation*}
\frac{1}{\beta _{(2K)}}=-\frac{1}{\beta _{(2K+1)}}=K+\frac{1}{\beta _{(0)}}%
\text{, \ \ \ \ \ }\frac{1}{\beta _{(-2K)}}=-\frac{1}{\beta _{(-2K+1)}}=-K+%
\frac{1}{\beta _{(0)}},
\end{equation*}%
then corresponding hydrodynamic \textit{lattices} contain the Kupershmidt
hydrodynamic chains (\textbf{\ref{fuk}})%
\begin{eqnarray*}
\bar{B}_{t}^{n} &=&\bar{B}_{x}^{n+1}+M\bar{B}^{0}\bar{B}_{x}^{n}+n\bar{B}^{n}%
\bar{B}_{x}^{0}\text{, \ \ \ }n=0,1,2,..., \\
&& \\
\tilde{B}_{t}^{n} &=&\tilde{B}_{x}^{n+1}-L\tilde{B}^{0}\tilde{B}_{x}^{n}+n%
\tilde{B}^{n}\tilde{B}_{x}^{0}\text{, \ \ \ }n=0,1,2,...,
\end{eqnarray*}%
which are exactly exclusive cases $\beta =1/N$ found by a dispersionless
limit in \textbf{\cite{Blaszak}} (see also \textbf{\cite{Manasa}}). Thus,
any hydrodynamic reduction and any solution of corresponding 2+1 quasilinear
equations (see below) can be recalculated for the above hydrodynamic chains
with arbitrary integer indexes $K_{1}\leftrightarrow K_{2}$.

\section{Ideal gas dynamics}

The first and the last equations in (\textbf{\ref{e}}) can be written in the
conservative form%
\begin{equation*}
B_{t}^{0}=\partial _{x}[B_{(\gamma )}^{1}+\frac{1+\beta \gamma }{2\beta }%
(B^{0})^{2}]\text{,\ \ \ \ \ \ \ }\partial _{t}\left[ \left( B_{(\gamma
)}^{N-1}\right) ^{\frac{1}{\beta (N-1+\gamma )}}\right] =\partial _{x}\left[ 
\frac{B^{0}}{\beta }\left( B_{(\gamma )}^{N-1}\right) ^{\frac{1}{\beta
(N-1+\gamma )}}\right] .
\end{equation*}%
The ideal gas dynamics written in physical variables (see, for instance, 
\textbf{\cite{Nutku}})%
\begin{equation}
u_{t}=\partial _{x}(\frac{u^{2}}{2}+\frac{\upsilon ^{\beta }}{\beta })\text{%
, \ \ \ \ \ \ \ }\upsilon _{t}=\partial _{x}(u\upsilon )  \label{gas}
\end{equation}%
is a two component hydrodynamic reduction (see details in \textbf{\cite%
{Kuper2}}) of the Kupershmidt hydrodynamic chain (\textbf{\ref{e}}), where $%
\gamma =0$, $B^{0}=\beta u$ and $B^{1}=\upsilon ^{\beta }$.

Transformations $\beta _{(k)}\rightarrow \beta _{(k+1)}$ described in the
two previous sections are compatible with any hydrodynamic reductions of the
Kupershmidt hydrodynamic chains (see, for instance, \textbf{\cite%
{Maks+algebr}}). In this paper we consider these transformations for the
ideal gas dynamics (\textbf{\ref{gas}})%
\begin{equation*}
\partial _{t_{(0)}^{1}}\upsilon _{(0)}=\partial
_{t_{(0)}^{0}}(u_{(0)}\upsilon _{(0)})\text{, \ \ \ \ \ \ }\partial
_{t_{(0)}^{1}}u_{(0)}=\partial _{t_{(0)}^{0}}\left( \frac{u_{(0)}^{2}}{2}+%
\frac{\upsilon _{(0)}^{\beta _{(0)}}}{\beta _{(0)}}\right) .
\end{equation*}%
Its commuting flow (see, for instance, \textbf{\cite{Nutku}})%
\begin{equation*}
\partial _{t_{(0)}^{-1}}\upsilon _{(0)}=\partial _{t_{(0)}^{0}}u_{(0)}\text{%
, \ \ \ \ \ \ }\partial _{t_{(0)}^{-1}}u_{(0)}=\partial _{t_{(0)}^{0}}\frac{%
\upsilon _{(0)}^{\beta _{(0)}-1}}{\beta _{(0)}-1},
\end{equation*}%
is known as the ``nonlinear elasticity equation''. Let us apply the
reciprocal transformation described in the previous section. This is nothing
but a transition from the Euler to the Lagrangian coordinates. So, the 
\textit{first iteration} is%
\begin{equation*}
\partial _{t_{(1)}^{-1}}\upsilon _{(1)}=\partial _{t_{(1)}^{0}}u_{(1)}\text{%
, \ \ \ \ \ \ \ \ \ \ \ \ \ \ \ \ \ \ \ \ \ \ \ \ \ \ \ \ \ }\partial
_{t_{(1)}^{1}}\upsilon _{(1)}=\partial _{t_{(1)}^{0}}(u_{(1)}\upsilon _{(1)})%
\text{,}
\end{equation*}%
\begin{equation*}
\partial _{t_{(1)}^{-1}}u_{(1)}=\partial _{t_{(1)}^{0}}\frac{\upsilon
_{(1)}^{\beta _{(1)}-1}}{\beta _{(1)}-1}\text{, \ \ \ \ \ \ \ \ \ \ \ \ \ \
\ \ \ \ \ \ \ \ \ \ \ \ }\partial _{t_{(1)}^{1}}u_{(1)}=\partial
_{t_{(1)}^{0}}\left( \frac{u_{(1)}^{2}}{2}+\frac{\upsilon _{(1)}^{\beta
_{(1)}}}{\beta _{(1)}}\right) ,
\end{equation*}%
where%
\begin{equation*}
\upsilon _{(1)}=\frac{1}{\upsilon _{(0)}}\text{, \ \ \ \ \ \ }%
u_{(1)}=-u_{(0)}\text{, \ \ \ \ \ \ }\beta _{(1)}=-\beta _{(0)}\text{, \ \ \
\ \ \ }t_{(1)}^{-1}=t_{(0)}^{1}\text{, \ \ \ \ \ \ }t_{(1)}^{1}=t_{(0)}^{-1}.
\end{equation*}%
According to the previous section, the \textit{second iteration} is%
\begin{equation*}
\partial _{t_{(2)}^{-1}}\upsilon _{(2)}=\partial _{t_{(2)}^{0}}u_{(2)}\text{%
, \ \ \ \ \ \ \ \ \ \ \ \ \ \ \ \ \ \ \ \ \ \ \ \ \ \ \ \ \ }\partial
_{t_{(2)}^{1}}\upsilon _{(2)}=\partial _{t_{(2)}^{0}}(u_{(2)}\upsilon _{(2)})%
\text{,}
\end{equation*}%
\begin{equation*}
\partial _{t_{(2)}^{-1}}u_{(2)}=\partial _{t_{(2)}^{0}}\frac{\upsilon
_{(2)}^{\beta _{(2)}-1}}{\beta _{(2)}-1}\text{, \ \ \ \ \ \ \ \ \ \ \ \ \ \
\ \ \ \ \ \ \ \ \ \ \ \ }\partial _{t_{(2)}^{1}}u_{(2)}=\partial
_{t_{(2)}^{0}}\left( \frac{u_{(2)}^{2}}{2}+\frac{\upsilon _{(2)}^{\beta
_{(2)}}}{\beta _{(2)}}\right) ,
\end{equation*}%
where%
\begin{equation*}
\upsilon _{(2)}=\upsilon _{(1)}^{\beta _{(1)}-1}\text{, \ \ \ }%
u_{(2)}=(\beta _{(1)}-1)u_{(1)}\text{, \ \ \ }\beta _{(2)}=\frac{\beta _{(1)}%
}{\beta _{(1)}-1}\text{, \ \ \ }t_{(2)}^{0}=t_{(1)}^{-1}\text{, \ \ \ }%
t_{(2)}^{-1}=t_{(1)}^{0}\text{, \ \ \ }t_{(2)}^{1}=t_{(1)}^{-2}.
\end{equation*}%
The ideal gas dynamics (\textbf{\ref{gas}}) and the nonlinear elasticity
equation in the Riemann invariants have the form%
\begin{eqnarray}
r_{t}^{1} &=&[r^{1}-\varepsilon (r^{1}+r^{2})]r_{x}^{1}\text{,\ \ \ \ \ \ \
\ \ \ \ \ \ \ \ \ \ \ \ \ \ \ \ \ \ }r_{y}^{1}=(r^{1}-r^{2})^{2\varepsilon
}r_{x}^{1},  \notag \\
&&  \label{rik} \\
r_{t}^{2} &=&[r^{2}-\varepsilon (r^{1}+r^{2})]r_{x}^{2}\text{, \ \ \ \ \ \ \
\ \ \ \ \ \ \ \ \ \ \ \ \ \ \ \ \ }r_{y}^{2}=-(r^{1}-r^{2})^{2\varepsilon
}r_{x}^{2},  \notag
\end{eqnarray}%
where%
\begin{equation*}
u=\frac{1-2\varepsilon }{2}(r^{1}+r^{2})\text{, \ \ \ \ \ \ }\upsilon
=\left( \frac{r^{1}-r^{2}}{4}\right) ^{1-2\varepsilon }\text{, \ \ \ \ \ \ \ 
}\beta =\frac{2}{1-2\varepsilon }.
\end{equation*}%
If $\varepsilon =1/2$ in such an exceptional case%
\begin{eqnarray*}
r_{t}^{1} &=&2[(r^{1}+r^{2})+(r^{1}-r^{2})\ln (r^{1}-r^{2})]r_{x}^{1}\text{,
\ \ \ \ \ \ \ \ \ \ \ \ \ \ \ \ \ \ }r_{y}^{1}=(r^{1}-r^{2})r_{x}^{1}, \\
&& \\
r_{t}^{2} &=&2[(r^{1}+r^{2})-(r^{1}-r^{2})\ln (r^{1}-r^{2})]r_{x}^{2}\text{,
\ \ \ \ \ \ \ \ \ \ \ \ \ \ \ \ \ \ }r_{y}^{2}=-(r^{1}-r^{2})r_{x}^{2},
\end{eqnarray*}%
where%
\begin{equation*}
u=2(r^{1}+r^{2})\text{, \ \ \ \ \ \ }\upsilon =2\ln (r^{1}-r^{2})\text{, \ \
\ \ \ \ \ }\beta =\pm \infty .
\end{equation*}%
Let us consider the above chain of transformations for the initial
Kupershmidt chain ($\beta =2$). In such a case $\varepsilon _{(0)}=0$. This
case is trivial. The corresponding gas dynamics is decomposed in a couple of
separate so-called the Euler-Monge-Riemann-Hopf equations%
\begin{equation*}
r_{t}^{1}=r^{1}r_{x}^{1}\text{, \ \ \ \ \ \ \ }r_{t}^{2}=r^{2}r_{x}^{2}.
\end{equation*}%
The first iteration yields $\beta _{(1)}=-2$, $\varepsilon _{(1)}=1$. This
is the first nontrivial, but still linearizable case. This is the
two-component linearly degenerate system (see \textbf{\cite{Nutku}}). The
second iteration yields $\beta _{(2)}=2/3$, $\varepsilon _{(2)}=-1$. This is
the two-component chromatography system (see \textbf{\cite{Nutku}}). The
corresponding hydrodynamic type system is also linearizable (see \textbf{%
\cite{Nutku}}). All other iterations yield $\varepsilon
_{(2K-1)}=-\varepsilon _{(2K)}=K$. Corresponding hydrodynamic type systems
are linearizable too (see \textbf{\cite{Nutku}}).

Let us consider the Gibbons--Tsarev system (\textbf{\ref{gt}}) written in
the form (see \textbf{\cite{Manasa}})%
\begin{equation*}
\partial _{i}q^{k}=\frac{q^{k}\partial _{i}B^{0}}{q^{i}-q^{k}}\text{, \ \ \
\ \ \ }\partial _{ik}B^{0}=\frac{(q^{i}+q^{k})\partial _{i}B^{0}\partial
_{k}B^{0}}{(q^{i}-q^{k})^{2}}\text{, \ \ \ \ \ }i\neq k.
\end{equation*}%
Following \textbf{\cite{Manasa}} let us restrict our attention on two
component hydrodynamic reductions such that%
\begin{equation*}
q^{1}+q^{2}=0.
\end{equation*}%
Then this Gibbons--Tsarev system can be integrated%
\begin{equation}
B^{0}=R_{1}(r^{1})+R_{2}(r^{2})\text{, \ \ \ \ \ }q^{1}=-q^{2}=\frac{1}{2}%
[R_{1}(r^{1})-R_{2}(r^{2})],  \label{sub}
\end{equation}%
where $R_{1}(r^{1})$ and $R_{2}(r^{2})$ are arbitrary functions. The
corresponding linear systems describing conservation law densities $h$ and
commuting flows $w^{k}$ in the Riemann invariants (see details in \textbf{%
\cite{Tsar}})%
\begin{equation*}
\frac{\partial ^{2}h}{\partial R_{1}\partial R_{2}}=\frac{\varepsilon }{%
R_{1}-R_{2}}\left( \frac{\partial h}{\partial R_{1}}-\frac{\partial h}{%
\partial R_{2}}\right) \text{, \ \ \ \ }\frac{\partial ^{2}W}{\partial
R_{1}\partial R_{2}}=-\frac{\varepsilon }{R_{1}-R_{2}}\left( \frac{\partial W%
}{\partial R_{1}}-\frac{\partial W}{\partial R_{2}}\right) \text{,}
\end{equation*}%
where $w^{1}=\partial W/\partial R_{1}$ and $w^{2}=\partial W/\partial R_{2}$
arise in the gas dynamics (see the first example in this section and \textbf{%
\cite{Maks+eps}}). These equations are nothing but famous the
Euler--Darboux--Poisson equations. A lot of new particular solutions also
can be found in \textbf{\cite{Popowicz}}; all possible transformations of
the first order are described in \textbf{\cite{Aks}}.

Indeed, the substitution (\textbf{\ref{sub}}) in the hydrodynamic type
system (\textbf{\ref{2}})%
\begin{equation*}
r_{t}^{1}=\left( q^{1}+\frac{B^{0}}{\beta }\right) r_{x}^{1}\text{, \ \ \ \
\ \ \ }r_{t}^{2}=\left( q^{2}+\frac{B^{0}}{\beta }\right) r_{x}^{2}
\end{equation*}%
leads to the ideal gas dynamics (\textbf{\ref{rik}}). Its first negative
commuting flow (see (\textbf{\ref{cona}}))%
\begin{equation*}
r_{t^{-1}}^{1}=\frac{(\mathbf{H}_{-1})^{\beta -1}}{q^{1}}r_{x}^{1}\text{, \
\ \ \ \ \ \ }r_{t^{-1}}^{2}=\frac{(\mathbf{H}_{-1})^{\beta -1}}{q^{2}}%
r_{x}^{2}
\end{equation*}%
leads to the nonlinear elasticity equation (\textbf{\ref{rik}}), where $%
\mathbf{H}_{-1}=[R_{1}(r^{1})+R_{2}(r^{2})]^{2/\beta }$ is a consequence of%
\begin{equation*}
\partial _{k}\ln \mathbf{H}_{-1}=\frac{\partial _{k}B^{0}}{\beta q^{k}},
\end{equation*}%
which is a result of the compatibility conditions $\partial
_{t}(r_{t^{-1}}^{1})=\partial _{t^{-1}}(r_{t}^{1})$.

\section{Special values of index $\protect\beta $}

The Kupershmidt hydrodynamic lattice determined by a special set of the
parameters $\beta =1/N$ (where $N$ is an integer) was derived by M. Blaszak
in \textbf{\cite{Blaszak}}. If $\beta =L/M$ (where $L$ and $M$ are
integers), then the function $F(w)$ in (\textbf{\ref{total}}) can be
integrated via elementary functions. Thus, generating functions of
conservation laws and commuting flows (see the section \textbf{7}) contain
logarithmic parts. For instance,%
\begin{eqnarray*}
\beta &=&-\frac{1}{2}\text{, \ \ \ \ \ \ }\partial _{t^{2}}p=\partial
_{t^{0}}[\ln p+\mathbf{H}_{0}p^{1/2}+(\mathbf{H}_{1}+\mathbf{H}_{0}^{2})p],
\\
&& \\
\beta &=&-1\text{,\ \ \ \ \ \ \ }\partial _{t^{1}}p=\partial _{t^{0}}(\ln p+%
\mathbf{H}_{0}p)\text{, \ \ \ \ \ \ \ \ \ \ \ \ \ \ \ \ \ \ \ \ \ \ \ \ \ }%
\partial _{\tau }\mathbf{H}_{-2}=-\partial _{t^{0}}\left( \frac{\mathbf{H}%
_{-2}}{p}+\ln (\mathbf{H}_{-1}/p)\right) , \\
&& \\
\beta &=&1\text{, \ \ \ \ \ \ \ \ \ \ }\partial _{t^{-1}}p=\partial
_{t^{0}}\ln (p/\mathbf{H}_{-1})\text{,\ \ \ \ \ \ \ \ \ \ \ \ \ \ \ \ \ \ \
\ \ \ \ \ \ \ \ \ \ \ \ }\partial _{\tau }\mathbf{H}_{0}=\partial
_{t^{0}}\ln p\text{,} \\
&& \\
\beta &=&1/2\text{, \ \ \ \ \ \ \ }\partial _{t^{-2}}p=\partial
_{t^{0}}\left( \ln (p/\mathbf{H}_{-1})-\frac{\mathbf{H}_{-2}}{\mathbf{H}_{-1}%
}(p/\mathbf{H}_{-1})^{1/2}\right) \text{, \ \ \ \ \ }\partial _{\tau }%
\mathbf{H}_{1}=\partial _{t^{0}}[\ln p+\mathbf{H}_{0}p^{-1/2}].\text{ \ \ \
\ \ \ \ }
\end{eqnarray*}

Let us consider \textit{infinitely many component reductions} of the Benney
hierarchy (see (\textbf{\ref{bm}})). The simplest example is the
dispersionless limit of BKP hierarchy (see, for instance, \textbf{\cite%
{Bogdan}}). The first higher commuting flow of the Benney hierarchy is%
\begin{equation*}
\partial _{t^{2}}A^{k}=\frac{1}{3}[kA^{k+n-1}\partial _{x}+n\partial
_{x}A^{k+n-1}]\frac{\partial \mathbf{H}_{3}}{\partial A^{n}},
\end{equation*}%
where $\mathbf{H}_{3}=A^{3}+3A^{0}A^{1}$. Let us split this hydrodynamic
chain%
\begin{equation*}
A_{t^{2}}^{k}=A_{x}^{k+2}+A^{0}A_{x}^{k}+kA^{k-1}A_{x}^{1}+(k+1)A^{k}A_{x}^{0}%
\text{, \ \ \ \ \ }k=0\text{, }1\text{, ...}
\end{equation*}%
on the \textit{even} and \textit{odd} sub-chains%
\begin{eqnarray*}
A_{t^{2}}^{2k}
&=&A_{x}^{2k+2}+A^{0}A_{x}^{2k}+2kA^{2k-1}A_{x}^{1}+(2k+1)A^{2k}A_{x}^{0}%
\text{, \ \ \ \ \ }k=0\text{, }1\text{, ...} \\
&& \\
A_{t^{2}}^{2k+1}
&=&A_{x}^{2k+3}+A^{0}A_{x}^{2k+1}+(2k+1)A^{2k}A_{x}^{1}+2(k+1)A^{2k+1}A_{x}^{0}%
\text{, \ \ \ \ \ }k=0\text{, }1\text{, ...}
\end{eqnarray*}%
It is easy to see, that the infinitely many component reduction $A^{2k+1}=0$%
, \ $k=0$, $1$, ... is compatible with the above chain%
\begin{equation*}
\partial _{t^{2}}\tilde{A}_{(2)}^{k}=\partial _{x}\tilde{A}_{(2)}^{k+1}+%
\tilde{A}_{(2)}^{0}\partial _{x}\tilde{A}_{(2)}^{k}+(2k+1)\tilde{A}%
_{(2)}^{k}\partial _{x}\tilde{A}_{(2)}^{0}\text{, \ \ \ \ \ }k=0\text{, }1%
\text{, ...,}
\end{equation*}%
where $\tilde{A}_{(2)}^{k}\equiv A^{2k}$. The Riemann mapping for the Benney
hierarchy%
\begin{equation}
\lambda =\mu +\frac{A^{0}}{\mu }+\frac{A^{1}}{\mu ^{2}}+\frac{A^{2}}{\mu ^{3}%
}+...  \label{rima}
\end{equation}%
(under the above constraint $A^{2k+1}=0$) reduces to%
\begin{equation*}
\lambda =\mu +\frac{\tilde{A}_{(2)}^{0}}{\mu }+\frac{\tilde{A}_{(2)}^{1}}{%
\mu ^{3}}+\frac{\tilde{A}_{(2)}^{2}}{\mu ^{5}}+...
\end{equation*}%
The inverse series to (\textbf{\ref{rima}})%
\begin{equation*}
\mu =\lambda -\frac{\mathbf{H}_{1}}{\lambda }-\frac{\mathbf{H}_{2}}{\lambda
^{2}}-\frac{\mathbf{H}_{3}}{\lambda ^{3}}-...
\end{equation*}%
reduced to%
\begin{equation*}
\mu =\lambda -\frac{\mathbf{\tilde{H}}_{1}^{(2)}}{\lambda }-\frac{\mathbf{%
\tilde{H}}_{2}^{(2)}}{\lambda ^{3}}-\frac{\mathbf{\tilde{H}}_{3}^{(2)}}{%
\lambda ^{5}}-...
\end{equation*}%
also means that $\tilde{A}_{(2)}^{k}$ and $\mathbf{\tilde{H}}_{1}^{(2)}$ no
longer depend on \textit{odd} ``times'' $t^{2k+1}$. Let us call \textit{even}
``times'' $t^{2k}=y^{k}$ (i.e. $x=t^{0}=y^{0}$). The corresponding Gibbons
equation is%
\begin{equation*}
\lambda _{y^{1}}-(\mu ^{2}+\tilde{A}_{(2)}^{0})\lambda _{x}=\frac{\partial
\lambda }{\partial \mu }\left[ \mu _{y^{1}}-\partial _{x}\left( \frac{\mu
^{3}}{3}+\tilde{A}_{(2)}^{0}\mu \right) \right] .
\end{equation*}%
The second example can be obtained in a similar way. The second higher
commuting flow of the Benney hierarchy is%
\begin{equation*}
\partial _{t^{3}}A^{k}=\frac{1}{4}[kA^{k+n-1}\partial _{x}+n\partial
_{x}A^{k+n-1}]\frac{\partial \mathbf{H}_{4}}{\partial A^{n}},
\end{equation*}%
where $\mathbf{H}_{4}=A^{4}+4A^{0}A^{2}+2(A^{1})^{2}+2(A^{0})^{3}$. Let us
split this hydrodynamic chain%
\begin{equation*}
A_{t^{3}}^{k}=A_{x}^{k+3}+2A^{0}A_{x}^{k+1}+A^{1}A_{x}^{k}+kA^{k-1}A_{x}^{2}+(k+1)A^{k}A_{x}^{1}+[(k+2)A^{k+1}+3kA^{k-1}A^{0}]A_{x}^{0}%
\text{,\ \ \ \ }k=0\text{, }1\text{, ...}
\end{equation*}%
on the \textit{three} sub-chains $A_{t^{3}}^{3k}=A_{x}^{3k+3}+...$, $%
A_{t^{3}}^{3k+1}=A_{x}^{3k+4}+...$, $A_{t^{3}}^{3k+2}=A_{x}^{3k+5}+$... Then
the infinitely many component reduction $A^{3k}=A^{3k+2}=0$, \ $k=0$, $1$,
... is compatible with the above chain%
\begin{equation*}
\partial _{t^{3}}\tilde{A}_{(3)}^{k}=\partial _{x}\tilde{A}_{(3)}^{k+1}+%
\tilde{A}_{(3)}^{0}\partial _{x}\tilde{A}_{(3)}^{k}+(3k+2)\tilde{A}%
_{(3)}^{k}\partial _{x}\tilde{A}_{(3)}^{0}\text{, \ \ \ \ \ }k=0\text{, }1%
\text{, ...,}
\end{equation*}%
where $\tilde{A}_{(3)}^{k}\equiv A^{3k+1}$. The Riemann mapping (\textbf{\ref%
{rima}}) reduces to%
\begin{equation*}
\lambda =\mu +\frac{\tilde{A}_{(3)}^{0}}{\mu ^{2}}+\frac{\tilde{A}_{(3)}^{1}%
}{\mu ^{5}}+\frac{\tilde{A}_{(3)}^{2}}{\mu ^{8}}+...
\end{equation*}%
The reduced inverse series%
\begin{equation*}
\mu =\lambda -\frac{\mathbf{\tilde{H}}_{0}^{(2)}}{\lambda ^{2}}-\frac{%
\mathbf{\tilde{H}}_{1}^{(2)}}{\lambda ^{5}}-\frac{\mathbf{\tilde{H}}%
_{2}^{(2)}}{\lambda ^{8}}-...
\end{equation*}%
also means that $\tilde{A}_{(3)}^{k}$ and $\mathbf{\tilde{H}}_{0}^{(3)}$ no
longer depend on ``times'' $t^{3k+1}$ and $t^{3k+2}$. Let us call ``times'' $%
t^{3k}=z^{k}$ (i.e. $x=t^{0}=z^{0}$). The corresponding Gibbons equation is%
\begin{equation*}
\lambda _{z^{1}}-(\mu ^{3}+\tilde{A}_{(3)}^{0})\lambda _{x}=\frac{\partial
\lambda }{\partial \mu }\left[ \mu _{z^{1}}-\partial _{x}\left( \frac{\mu
^{4}}{4}+\tilde{A}_{(3)}^{0}\mu \right) \right] .
\end{equation*}%
Thus, it is easy to generalize the above examples. Let us introduce $%
\tilde{A}_{(N)}^{k}\equiv A^{N(k+1)-2}$, where $N$ is an arbitrary natural
number. All other moments vanish. Such infinitely many component reduction
is connected with the Riemann mapping%
\begin{equation*}
\lambda =\mu +\frac{\tilde{A}_{(N)}^{0}}{\mu ^{N-1}}+\frac{\tilde{A}%
_{(N)}^{1}}{\mu ^{2N-1}}+\frac{\tilde{A}_{(N)}^{2}}{\mu ^{3N-1}}+...
\end{equation*}%
The corresponding hydrodynamic chain%
\begin{equation*}
\partial _{t^{N}}\tilde{A}_{(N)}^{k}=\partial _{x}\tilde{A}_{(N)}^{k+1}+%
\tilde{A}_{(N)}^{0}\partial _{x}\tilde{A}_{(N)}^{k}+[(k+1)N-1]\tilde{A}%
_{(N)}^{k}\partial _{x}\tilde{A}_{(N)}^{0}\text{, \ \ \ \ }k=0\text{, }1%
\text{, }2\text{, ...}
\end{equation*}%
satisfies the Gibbons equation%
\begin{equation*}
\lambda _{\tau ^{1}}-(\mu ^{N}+\tilde{A}_{(N)}^{0})\lambda _{x}=\frac{%
\partial \lambda }{\partial \mu }\left[ \mu _{\tau ^{1}}-\partial _{x}\left( 
\frac{\mu ^{N+1}}{N+1}+\tilde{A}_{(N)}^{0}\mu \right) \right] .
\end{equation*}%
All moments $\tilde{A}_{(N)}^{k}$ are functions of `times'' $\tau ^{k}\equiv
t^{kN}$, $k=0$, $1$, $2$, ... The substitution $B_{(1-1/N)}^{k}\equiv N%
\tilde{A}_{(N)}^{k}$ yields the Kupershmidt hydrodynamic chain (\textbf{\ref%
{fuk}}). Thus, infinitely many discrete values $\beta =N$ can be obtained
directly from the Benney hierarchy (\textbf{\ref{bm}}) by the above
described degeneration.

\section{Quasilinear equations and their particular solutions}

The extended Kupershmidt lattice is connected with an infinite set of 2+1
quasilinear equations, which can be obtained by an appropriate elimination
of the conservation law densities $\mathbf{H}_{k}$ and auxiliary time
variables $t^{n}$. Without lost of generality let us restrict our
consideration for simplicity on two first 2+1 quasilinear equations (see
generating functions of conservation laws and commuting flows in the section 
\textbf{7})%
\begin{eqnarray*}
\partial _{\tau }\mathbf{H}_{0} &=&\partial _{t^{0}}\frac{p^{\beta -1}}{%
\beta -1}\text{, \ \ \ \ \ \ \ \ \ \ \ \ \ \ \ \ \ \ \ \ \ \ \ \ \ \ \ \ \ \
\ \ \ \ \ \ \ \ \ \ \ \ \ \ \ \ \ \ \ }\partial _{\tau }\mathbf{H}%
_{-1}=-\partial _{t^{0}}\frac{\mathbf{H}_{-1}}{p}\text{,} \\
&& \\
\partial _{t^{1}}p &=&\partial _{t^{0}}\left( \frac{p^{\beta +1}}{\beta +1}+%
\mathbf{H}_{0}p\right) \text{, \ \ \ \ \ \ \ \ \ \ \ \ \ \ \ \ \ \ \ \ \ \ \ 
}\partial _{t^{-1}}p=\partial _{t^{0}}\left( \frac{(p/\mathbf{H}%
_{-1})^{1-\beta }}{1-\beta }\right) .
\end{eqnarray*}%
Substituting the asymptotics (\textbf{\ref{ful}}) one can obtain infinitely
many 2+1 quasilinear equations, where the first of them are%
\begin{eqnarray*}
\Phi _{xy} &=&\Phi _{tt}+(\beta -1)\Phi _{x}\Phi _{xt}+(\Phi _{t}-\beta \Phi
_{x}^{2}/2)\Phi _{xx}, \\
&& \\
\Phi _{\sigma t} &=&(\beta -1)\Phi _{\sigma }\Phi _{xx}+\Phi _{x}\Phi
_{\sigma x},
\end{eqnarray*}%
where $\mathbf{H}_{0}=\Phi _{x}$, $\mathbf{H}_{1}=\Phi _{t}+(\beta -2)\Phi
_{x}^{2}/2$, $(\mathbf{H}_{-1})^{\beta -1}=(\beta -1)\Phi _{\sigma }$ and $%
y=t^{2}$, $\sigma =t^{-1}$. These 2+1 quasilinear equations (for $\beta =1/N$
) were derived in \textbf{\cite{Blaszak}}; the first of them is the so
called ``$N-$dmKP'', the second equation is the so called ``$N-$dDym''.
These 2+1 quasilinear equations also can be derived from the compatibility
conditions $\partial _{t^{1}}(\partial _{t^{2}}p)=\partial _{t^{2}}(\partial
_{t^{1}}p)$ and $\partial _{t^{1}}(\partial _{t^{-1}}p)=\partial
_{t^{-1}}(\partial _{t^{1}}p)$. The third equation (for $\beta =1/N$ is the
so called ``$N-$dToda'')%
\begin{equation*}
(\beta -1)^{1/(\beta -1)}\Phi _{\sigma \sigma }=(\Phi _{\sigma })^{-1/(\beta
-1)}\Phi _{xz}-\Phi _{z}(\Phi _{\sigma })^{-\beta /(\beta -1)}\Phi _{\sigma
x},
\end{equation*}%
where $\mathbf{H}_{-2}=(\beta -1)^{(2-\beta )/(\beta -1)}\Phi _{z}(\Phi
_{\sigma })^{(2-\beta )/(\beta -1)}$ and$\ z=t^{-2}$ was derived in \textbf{%
\cite{Manasa}} and can be obtained from the compatibility condition $%
\partial _{t^{-1}}(\partial _{t^{-2}}p)=\partial _{t^{-2}}(\partial
_{t^{-1}}p)$. However, this third 2+1 quasilinear equation is equivalent the
second 2+1 quasilinear equation (``$N-$dDym'') up to the transformation $%
t^{k}\leftrightarrow \tilde{t}^{-1-k}$ described in the section \textbf{7}.

The generalized hodograph method (see \textbf{\cite{Tsar}}) allows to
construct a general solution for $N$ component hydrodynamic reductions
described by the Gibbons--Tsarev system (\textbf{\ref{gt}}). The
Gibbons--Tsarev system has a general solution parameterized by $N$ arbitrary
functions of a single variable. We are not able to find this general
solution at this moment. However, infinitely many particular solutions
(parameterized by the hypergeometric function) are already known (see the
section \textbf{4}). Since the generating function of commuting flows (see (%
\textbf{\ref{total}})) is found for the Kupershmidt hydrodynamic chain and
for all its hydrodynamic reductions, then particular solutions for the above
2+1 quasilinear equations can be found by the generalized hodograph method
written in the Riemann coordinates $r^{k}$ (see (\textbf{\ref{2}}))%
\begin{equation*}
x+t\upsilon _{(1)}^{i}(\mathbf{r})+y\upsilon _{(2)}^{i}(\mathbf{r}%
)+z\upsilon _{(3)}^{i}(\mathbf{r})=w^{i}(\mathbf{r})
\end{equation*}%
or via field variables (conservation law densities, see (\textbf{\ref{3}})) $%
b^{k}$%
\begin{equation}
x\delta _{k}^{i}+t\upsilon _{(1)k}^{i}(\mathbf{b})+y\upsilon _{(2)k}^{i}(%
\mathbf{b})+z\upsilon _{(3)k}^{i}(\mathbf{b})=w_{k}^{i}(\mathbf{b}),
\label{gh}
\end{equation}%
where the corresponding hydrodynamic reductions%
\begin{equation*}
r_{t^{n}}^{i}=\upsilon _{(n)}^{i}(\mathbf{r})r_{t^{0}}^{i}\text{ \ \ \ \ \ \ 
}\Leftrightarrow \text{ \ \ \ \ \ \ \ }b_{t^{n}}^{i}=\upsilon _{(n)k}^{i}(%
\mathbf{b})b_{t^{0}}^{k}\text{, \ \ \ \ }i,k=1,2,...,N\text{, \ \ \ \ }n=\pm
1,\pm 2
\end{equation*}%
are written below in the explicit form in the Riemann invariants%
\begin{eqnarray*}
r_{t}^{k} &=&[(r^{k})^{\beta }+\mathbf{h}_{0}]r_{x}^{k}\text{, \ \ \ }%
r_{y}^{k}=[(r^{k})^{2\beta }+(\beta +1)\mathbf{h}_{0}(r^{k})^{\beta }+%
\mathbf{h}_{1}+(\mathbf{h}_{0})^{2}]r_{x}^{k}\text{,} \\
&& \\
r_{\sigma }^{k} &=&(\mathbf{h}_{-1})^{\beta -1}(r^{k})^{-\beta }r_{x}^{k}%
\text{, \ \ \ \ \ }r_{z}^{k}=[(\mathbf{h}_{-1})^{2\beta -1}(r^{k})^{-2\beta
}+(\beta -1)\mathbf{h}_{-2}(\mathbf{h}_{-1})^{\beta -2}(r^{k})^{1-\beta
}]r_{x}^{k}\text{,}
\end{eqnarray*}%
and in the conservative form%
\begin{eqnarray*}
b_{t}^{i} &=&\partial _{x}\left( \frac{(b^{i})^{\beta +1}}{\beta +1}+\mathbf{%
h}_{0}b^{i}\right) \text{, \ \ \ }b_{y}^{i}=\partial _{x}\left( \frac{%
(b^{i})^{2\beta +1}}{2\beta +1}+\mathbf{h}_{0}(b^{i})^{\beta +1}+(\mathbf{h}%
_{1}+(\mathbf{h}_{0})^{2})b^{i}\right) \text{,} \\
&& \\
b_{\sigma }^{i} &=&\partial _{x}\left( \frac{(b^{i}/\mathbf{h}%
_{-1})^{1-\beta }}{1-\beta }\right) \text{, \ \ \ \ \ \ \ \ \ }%
b_{z}^{i}=\partial _{x}\left( \frac{(b^{i}/\mathbf{h}_{-1})^{1-2\beta }}{%
1-2\beta }-\mathbf{h}_{-2}(\mathbf{h}_{-1})^{\beta -2}(b^{i})^{1-\beta
}\right) \text{.}
\end{eqnarray*}

Without lost of generality let us restrict\ our consideration for simplicity
on $N$ component hydrodynamic reductions determined by the moment
decomposition (\textbf{\ref{one}}). Then $N$ infinite series of the
conservation law densities $h_{n}^{(k)}$ can be found by the B\"{u}%
rmann--Lagrange expansion (\textbf{\ref{giper}}) from the equation of the
Riemann surface%
\begin{equation*}
\lambda =[1-(\frac{b^{i}}{p^{(i)}})^{\beta }]\exp \left[ \underset{n=1}{%
\overset{\infty }{\sum }}\frac{(-1)^{n}(\gamma -1)!}{nn!(\gamma -1-n)!}[1-(%
\frac{b^{i}}{p^{(i)}})^{\beta }]^{n}-\frac{1}{\varepsilon ^{i}}\underset{%
m\neq i}{\sum }\varepsilon _{m}\underset{n=0}{\overset{\infty }{\sum }}\frac{%
(b^{m}/p^{(i)})^{\beta (n+\gamma )}}{n+\gamma }-\frac{\beta (p^{(i)})^{\beta
(1-\gamma )}}{\varepsilon ^{i}(1-\gamma )}\right]
\end{equation*}%
at the vicinity of the every puncture $p_{0}^{(i)}=b^{i}$:%
\begin{eqnarray*}
h_{1}^{(i)} &=&\frac{b^{i}}{\beta }\exp \left[ \frac{1}{\varepsilon _{i}}%
\left( \frac{\beta (b^{i})^{\beta (1-\gamma )}}{1-\gamma }+\underset{m\neq i}%
{\sum }\varepsilon _{m}\left( \frac{b^{m}}{b^{i}}\right) ^{\beta \gamma
}F\left( 1,\gamma ,\gamma +1,\left( \frac{b^{m}}{b^{i}}\right) ^{\beta
}\right) \right) \right] \text{,} \\
&& \\
h_{2}^{(i)} &=&\frac{(h_{1}^{(i)})^{2}}{b^{i}}\left[ \frac{1-\beta }{2}%
+\beta \gamma +\frac{\beta (b^{i})^{\beta (1-\gamma )}}{\varepsilon _{i}}%
\left( \beta -\underset{m\neq i}{\sum }\frac{\varepsilon _{m}(b^{m})^{\beta
\gamma }}{(b^{i})^{\beta }-(b^{m})^{\beta }}\right) \right] ,...
\end{eqnarray*}

Since the generating function of commuting flows (replace $p\rightarrow
b^{i} $ in (\textbf{\ref{total}})) is%
\begin{equation*}
b_{\tau (\zeta )}^{i}=-\frac{1}{\beta }\partial _{x}\left[ \frac{b^{i}}{%
p(\zeta )}F\left( 1,\frac{1}{\beta },\frac{\beta +1}{\beta },\left( \frac{%
b^{i}}{p(\zeta )}\right) ^{\beta }\right) \right] ,
\end{equation*}%
then the Taylor expansion at the vicinity $p_{0}^{(k)}=b^{k}$ (see \textbf{%
\cite{Maks+algebr}}) yields $N$ infinite series of commuting flows%
\begin{equation*}
b_{t^{k,n}}^{i}=\partial _{x}w_{(k,n)}^{i}(\mathbf{b})\text{, \ \ \ \ }%
i,k=1,2,...,N\text{, \ \ \ \ \ }n=0,1,2,...
\end{equation*}%
For instance, the first such commuting flow is%
\begin{eqnarray*}
b_{t^{i,0}}^{k} &=&-\frac{1}{\beta }\partial _{x}\left[ \frac{b^{k}}{b^{i}}%
F\left( 1,\frac{1}{\beta },\frac{\beta +1}{\beta },\left( \frac{b^{k}}{b^{i}}%
\right) ^{\beta }\right) \right] \text{, \ \ \ \ }k\neq i, \\
&& \\
b_{t^{i,0}}^{i} &=&\frac{1}{\varepsilon _{i}}\partial _{x}\left[ \frac{%
(b^{i})^{1-\gamma }}{1-\gamma }-\frac{1}{\beta }\underset{n\neq i}{\sum }%
\varepsilon _{n}\left( \frac{b^{n}}{b^{i}}\right) ^{\beta \gamma }F\left(
1,\gamma ,\gamma +1,\left( \frac{b^{n}}{b^{i}}\right) ^{\beta }\right) %
\right] .
\end{eqnarray*}%
Thus, the generalized hodograph method (\textbf{\ref{gh}}) yields infinitely
many particular solutions%
\begin{equation*}
x\delta _{k}^{i}+t\partial _{k}\left( \frac{(b^{i})^{\beta +1}}{\beta +1}+%
\mathbf{h}_{0}b^{i}\right) +y\partial _{k}\left( \frac{(b^{i})^{2\beta +1}}{%
2\beta +1}+\mathbf{h}_{0}(b^{i})^{\beta +1}+(\mathbf{h}_{1}+(\mathbf{h}%
_{0})^{2})b^{i}\right)
\end{equation*}%
\begin{equation*}
+\sigma \partial _{k}\left( \frac{(b^{i}/\mathbf{h}_{-1})^{1-\beta }}{%
1-\beta }\right) +z\partial _{k}\left( \frac{(b^{i}/\mathbf{h}%
_{-1})^{1-2\beta }}{1-2\beta }-\mathbf{h}_{-2}(\mathbf{h}_{-1})^{\beta
-2}(b^{i})^{1-\beta }\right) =\partial _{k}w_{(m,n)}^{i}.
\end{equation*}

\section{Conclusion and outlook}

Solutions of 2+1 quasilinear equations and corresponding hydrodynamic chains
depend on integrability of the Gibbons--Tsarev system (\textbf{\ref{gt}}).
In this paper we present multi-parametric family of hydrodynamic reductions,
whose Riemann surfaces are associated with the hypergeometric function $%
_{2}F_{1}(1,\gamma ,\gamma +1,z)$. Taking into account another class of
hydrodynamic reductions derived in \textbf{\cite{Gib+Yu}} for the Benney
hydrodynamic chain (\textbf{\ref{bm}}), whose Riemann surfaces are
associated with hyperelliptic integrals, we believe that more general and
complicated hydrodynamic reductions can be found.

However, the main result of this paper is the existence of an infinite set
of local Hamiltonian structures. Since the sub-case $\beta =1$ is connected
with the Benney hydrodynamic chain (\textbf{\ref{bm}}) by the Miura type
transformation, it means that the Benney hydrodynamic chain also has an
infinite series of local Hamiltonian structures.

The Kupershmidt hydrodynamic lattice has four very interesting and important
sub-cases: $\beta =1$, $\beta =2$, $\beta =\infty $ and $\beta =0$. The
first sub-case is the modified dKP hierarchy, the second sub-case is the
dBKP hierarchy, the third case is a continuum limit of the 2DToda hierarchy,
the fourth case we call the universal hierarchy (see \textbf{\cite{Alonso}}
and \textbf{\cite{Maks+eps}}). Just the sub-case $\beta =2$ can be easily
extracted from this paper. In all other sub-cases all formulas have
singularities for corresponding values of the index $\beta $. These cases $%
\beta =1$, $\beta =\infty $ (and $\beta =2$) determine the Egorov
hydrodynamic chains connected by the aforementioned transformations and by
the Miura type transformation. They are considered in details in \textbf{%
\cite{Maks+wdvv}}.

\section*{Acknowledgement}

I thank Maciej Blaszak, Eugeni Ferapontov, John Gibbons, Yuji Kodama, Boris
Konopelchenko, Boris Kupershmidt, Andrey Maltsev and Sergey Tsarev for their
help and clarifying discussions.

I am grateful to the Institute of Mathematics in Taipei (Taiwan) where some
part of this work has been done, and especially to Jen-Hsu Chang, Jyh-Hao
Lee, Ming-Hien Tu and Derchyi Wu for fruitful discussions.%


\addcontentsline{toc}{section}{References}

\begin{thebibliography}{99}
\bibitem{Aks} \emph{A.V. Aksenov}, \newblock Symmetries and relations
between solutions of a class of Euler-Poisson-Darboux equations. (Russian)
Dokl. Akad. Nauk. (Reports of RAS), \textbf{381} No. 2 (2001) 176--179.

\bibitem{Alonso} \emph{L.M. Alonso, A.B. Shabat}, \newblock Energy-dependent
potentials revisited: A universal hierarchy of hydrodynamic type, Phys.
lett. A, \textbf{299} No. 4 (2002) 359-365.

\bibitem{Benney} \emph{D.J. Benney}, \newblock Some properties of long
non-linear waves, Stud. Appl. Math., \textbf{52} (1973) 45-50.

\bibitem{Blaszak} \emph{M. Blaszak}, \newblock Classical $R-$matrices on
Poisson algebras and related dispersionless systems Phys. Lett. A, \textbf{%
297} (2002), 191-195. \emph{M. B\l aszak, B.M. Szablikowski}, \newblock %
Classical $R-$matrix theory of dispersionless systems: I. (1+1)-dimension
theory, J. Phys. A: Math. Gen., \textbf{35} (2002) 10325-10344. \emph{M. B\l
aszak, B.M. Szablikowski}, \newblock Classical $R-$matrix theory of
dispersionless systems: II. (2+1)-dimension theory, J. Phys. A: Math. Gen., 
\textbf{35} (2002) 10345-10364. \emph{M. B\l aszak, B.M. Szablikowski}, %
\newblock Meromorphic Lax representations of (1+1)-dimensional
multi-Hamiltonian dispersionless systems, submitted for publication, arXiv:
nlin.SI/0510068. \emph{B.M. Szablikowski}, \newblock Gauge transformation
and reciprocal link for (2+1)-dimensional integrable field systems, J.
Nonlinear Math. Phys., \textbf{13} (2006) 117-128.

\bibitem{Bogdan} \emph{L.V. Bogdanov, B.G. Konopelchenko}, \newblock %
Symmetry constraints for dispersionless integrable equations and systems of
hydrodynamic type, Phys. Lett. A, \textbf{330 }(2004) 448--459.

\bibitem{Dorfman} \emph{I.Ya. Dorfman}, \newblock Dirac structures and
integrability of nonlinear evolution equations; Nonlinear Science: Theory
and Applications, John Wiley \& Sons, New York (1993) 176 pp.

\bibitem{Dubr+Nov} \emph{B.A. Dubrovin, S.P. Novikov,} \newblock Hamiltonian
formalism of one-dimensional systems of hydrodynamic type and the
Bogolyubov-Whitham averaging method, Soviet Math. Dokl., \textbf{27} (1983)
665--669. \emph{B.A. Dubrovin, S.P. Novikov,} \newblock Hydrodynamics of
weakly deformed soliton lattices. Differential geometry and Hamiltonian
theory, Russian Math. Surveys, \textbf{44} No. 6 (1989) 35--124.

\bibitem{Fer+trans} \emph{E.V. Ferapontov,} \newblock Nonlocal Hamiltonian
operators of hydrodynamic type: differential geometry and applications,
Amer. Math. Soc. Transl. (2), \textbf{170} (1995) 33-58.

\bibitem{Fer+Dav} \emph{E.V. Ferapontov, D.G. Marshall}, \newblock %
Differential-geometric approach to the integrability of hydrodynamic chains:
the Haanties tensor, \textit{arXiv:nlin.SI/0505013}.

\bibitem{Fer+Mokh} \emph{E.V. Ferapontov, O.I. Mokhov}, \newblock Nonlocal
Hamiltonian operators of hydrodynamic type that are connected with metrics
of constant curvature, Russian Math. Surveys, \textbf{45} No. 3 (1990)
218--219.

\bibitem{Gibbons} \emph{J. Gibbons}, \newblock Collisionless Boltzmann
equations and integrable moment equations, Physica D, \textbf{3} (1981)
503-511.

\bibitem{Gib+Tsar} \emph{J. Gibbons, S.P. Tsarev}, \newblock Reductions of
Benney's equations, Phys. Lett. A, \textbf{211 }(1996) 19-24. \emph{J.
Gibbons, S.P. Tsarev}, \newblock Conformal maps and reductions of the Benney
equations, Phys. Lett. A, \textbf{258} (1999) 263-270.

\bibitem{Gib+Yu} \emph{J. Gibbons, L.A. Yu}, \newblock The initial value
problem for reductions of the Benney equations, Inverse Problems, \textbf{16}
No. 3 (2000) 605-618, \emph{L.A. Yu}, \newblock Waterbag reductions of the
dispersionless discrete KP hierarchy, J. Phys. A: Math. Gen., \textbf{33}
(2000) 8127--8138.

\bibitem{Kuper} \emph{B.A. Kupershmidt}, \newblock Deformations of
integrable systems, Proc. Roy. Irish Acad. Sect. A, \textbf{83} No. 1 (1983)
45-74. \emph{B.A. Kupershmidt}, \newblock Normal and universal forms in
integrable hydrodynamical systems, Proceedings of the Berkeley-Ames
conference on nonlinear problems in control and fluid dynamics (Berkeley,
Calif., 1983), in Lie Groups: Hist., Frontiers and Appl. Ser. B: Systems
Inform. Control, II, Math Sci Press, Brookline, MA, (1984) 357-378.

\bibitem{KM} \emph{B.A. Kupershmidt, Yu.I. Manin}, \newblock Long wave
equations with a free surface. II. The Hamiltonian structure and the higher
equations, Func. Anal. Appl., \textbf{12} No. 1 (1978) 25--37. \emph{D.R.
Lebedev, Yu.I. Manin}, \newblock Conservation laws and representation of
Benney's long wave equations, Phys. Lett. A, \textbf{74} No. 3,4 (1979)
154-156.

\bibitem{Kuper1} \emph{B.A. Kupershmidt}, \newblock Hydrodynamic chains of
Pavlov's class, accepted in Phys. Lett. A.

\bibitem{Kuper2} \emph{B.A. Kupershmidt}, \newblock Equations of long waves
with a free surface III. The multidimensional case, J. Nonlinear Math.
Phys., \textbf{12} No. 4 (2005) 539-549. \emph{B.A. Kupershmidt}, \newblock %
Extensions of 1-dimensional polytropic gas dynamics, J. Nonlinear Math.
Phys., \textbf{13} No. 1 (2006) 145-157.

\bibitem{Lavr} \emph{M.A. Lavrentiev, B.V. Shabat}, 
\newblock {Metody teorii
funktsi\u\i kompleksnogo peremennogo}(Russian) [Methods of the theory of
functions of a complex variable] Third corrected edition Izdat. ``Nauka'',
Moscow (1965) 716 pp.

\bibitem{Malt+Nov} \emph{A.Ya. Maltsev, S.P. Novikov}, \newblock On the
local systems Hamiltonian in the weakly nonlocal Poisson brackets, Physica
D, \textbf{156} (2001) 53-80.

\bibitem{Manas} \emph{M. Manas, L.M. Alonso, E. Medina}, \newblock %
Reductions of the dispersionless KP hierarchy, Theor. Math. Phys., \textbf{%
133} No. 3 (2002) 1712-1721, \emph{B. Konopelchenko, L.M. Alonso, E. Medina}%
, \newblock Quasiconformal mappings and solutions of the dispersionless KP
hierarchy, Theor. Math. Phys., \textbf{133} No. 2 (2002) 1529-1538.

\bibitem{Manasa} \emph{M. Manas}, \newblock$S-$functions, reductions and
hodograph solutions of the $r$th dispersionless modified KP and Dym
hierarchies, J. Phys. A: Math. Gen., \textbf{37} (2004) 11191--11221. \emph{%
M. Manas}, \newblock On the $r$th dispersionless Toda hierarchy:
factorization problem, additional symmetries and some solutions, J. Phys. A:
Math. Gen., \textbf{37} (2004) 9195-9224.

\bibitem{Nutku} \emph{Y. Nutku}, \newblock On a new class of completely
integrable nonlinear wave equations. Multi-Hamiltonian structure II, J.
Math. Phys., \textbf{28} No. 11 (1987) 2579--2585.

\bibitem{Maks+Egor} \emph{M.V. Pavlov}, \newblock Classification of the
Egorov hydrodynamic chains, Theor. Math. Phys., \textbf{138} No. 1 (2004)
55-71.

\bibitem{Maks+eps} \emph{M.V. Pavlov}, \newblock Integrable hydrodynamic
chains, J. Math. Phys., \textbf{44} No. 9 (2003) 4134-4156.

\bibitem{Maks+cc} \emph{M.V. Pavlov}, \newblock Integrable systems and
metrics of constant curvature, J. Nonlinear Math. Phys., No. 9 Supplement 1
(2002) 173-191.

\bibitem{Maks+Tsar} \emph{M. V. Pavlov, S.P. Tsarev,} \newblock
Three-Hamiltonian structures of the Egorov hydrodynamic type systems, Funct.
Anal. Appl., \textbf{37} No. 1 (2003) 32-45.

\bibitem{Maks+Puas} \emph{M.V. Pavlov}, \newblock Hydrodynamic chains and
the classification of their Poisson brackets.

\bibitem{Maks+algebr} \emph{M.V. Pavlov}, \newblock Algebro-geometric
approach in the theory of integrable hydrodynamic type systems.

\bibitem{Maks+Hamch} \emph{M.V. Pavlov}, \newblock The Hamiltonian approach
in the classification and the integrability of hydrodynamic chains.

\bibitem{Maks+vech} \emph{M.V. Pavlov}, \newblock Classification of
integrable hydrodynamic chains and generating functions of conservation laws.

\bibitem{Maks+wdvv} \emph{M.V. Pavlov}, \newblock Explicit solutions of the
WDVV equation determined by the ``flat'' hydrodynamic reductions of the
Egorov hydrodynamic chains.

\bibitem{Popowicz} \emph{M.V. Pavlov, Z. Popowicz}, \newblock Non-polynomial
conservation law densities generated by the symmetry operators in some
hydrodynamical models. J. Phys. A.: Math. and Gen., \textbf{36} (2003) 1-10.

\bibitem{Rogers} \emph{C. Rogers, W.~F. Shadwick}, \newblock B\"{a}cklund
Transformations and their Applications, Academic Press (1982) NewYork.

\bibitem{Yanenko} \emph{B.L. Rozhdestvenski, N.N. Yanenko}, \newblock %
Systems of quasilinear equations and their applications to gas dynamics.
Translated from the second Russian edition by J. R. Schulenberger.
Translations of Mathematical Monographs, 55. American Mathematical Society,
Providence, RI, 1983; Russian ed. Nauka, (1968) Moscow.

\bibitem{Tsar} \emph{S.P. Tsarev}, \newblock On Poisson brackets and
one-dimensional Hamiltonian systems of hydrodynamic type, Soviet Math.
Dokl., \textbf{31} (1985) 488--491. \emph{S.P. Tsarev}, \newblock The
geometry of Hamiltonian systems of hydrodynamic type. The generalized
hodograph method, Math. USSR Izvestiya, \textbf{37} No. 2 (1991) 397--419.

\bibitem{Zakh} \emph{V.E. Zakharov}, \newblock Benney's equations and
quasi-classical approximation in the inverse problem method, Funct. Anal.
Appl., \textbf{14} No. 2 (1980) 89-98. \emph{V.E. Zakharov}, \newblock On
the Benney's Equations, Physica 3D, (1981) 193-200.
\end{thebibliography}
\end{document}